\documentclass[preprint]{aastex63}
\usepackage[utf8]{inputenc}

\begin{document}

\title{GJ 229B: Solving the Puzzle of the First Known T-Dwarf with the APOLLO Retrieval Code}

\author[0000-0002-4884-7150]{Alex R. Howe}
\affiliation{NASA Goddard Space Flight Center, 8800 Greenbelt Rd, Greenbelt, MD 20771, USA}

\author[0000-0003-0241-8956]{Michael W. McElwain}
\affiliation{NASA Goddard Space Flight Center, 8800 Greenbelt Rd, Greenbelt, MD 20771, USA}

\author[0000-0002-8119-3355 ]{Avi M. Mandell}
\affiliation{NASA Goddard Space Flight Center, 8800 Greenbelt Rd, Greenbelt, MD 20771, USA}
\affiliation{GSFC Sellers Exoplanet Environments Collaboration}

\submitjournal{AAS Journals}

\begin{abstract}

GJ 229B was the first T-dwarf to be discovered in 1995, and its spectrum has been more thoroughly observed than most other brown dwarfs. Yet a full spectroscopic analysis of its atmosphere has not been done with modern techniques. This spectrum has several peculiar features, and recent dynamical estimates of GJ 229B's mass and orbit have disagreed widely, both of which warrant closer investigation. With a separation of tens of AU from its host star, GJ 229B falls near the border of the planet and stellar population formation regimes, so its atmosphere could provide clues to formation processes for intermediate objects of this type. In an effort to resolve these questions, we performed retrievals on published spectra of GJ 229B over a wide range of wavelengths (0.5-5.1 ${\rm \mu m}$) using the open-source APOLLO code. Based on these retrievals, we present a more precise mass estimate of $41.6\pm3.3\, M_J$ and an effective temperature estimate of $869_{-7}^{+5}$ K, which are more consistent with evolutionary models for brown dwarfs and suggest an older age for the system of $>$1.0 Gyr. We also present retrieved molecular abundances for the atmosphere, including replicating the previously-observed high CO abundance, and discuss their implications for the formation and evolution of this object. This retrieval effort will give us insight into how to study other brown dwarfs and directly-imaged planets, including those observed with {\it JWST} and other next-generation telescopes.

\end{abstract}

\section{Introduction}

GJ 229B was the first type-T brown dwarf to be discovered \citep{GJ229Disc}\footnote{Not to be confused with the Neptune-mass planet accompanying the host star, GJ 229Ab, sometimes written as GJ 229b \citep{GJ229Planet}.}, typified by observable quantities of methane in its atmosphere and an effective temperature well below the L/T transition at $\sim$1200 K. With an initial projected separation from its host star of 7.8 arcsec, this object is a prime candidate for direct imaging with modern spectroscopic techniques. Moreover, we now know that brown dwarfs like GJ 229B bear many similarities to directly-imaged giant exoplanets found orbiting young stars (e.g. \citealt{HR8799Disc}). Thus, further study of such a well-documented object could greatly inform our understanding of giant planetary companions.

In the years following its discovery, several spectra were taken of GJ 229B \citep{GJ229UKblue,GJ229Keck,GJ229UKred,GJ229HST}. These spectra span a wavelength range from the visible at 0.52 ${\rm \mu m}$ into the mid-infrared at 5.1 ${\rm \mu m}$, a wider range than what is available for many similar objects and comparable to the higher-quality spectra that will be available from the \textit{James Webb Space Telescope (JWST)}. Despite this, a full spectroscopic retrieval of GJ 229B's atmosphere has only recently been attempted \citep{GJ229BRet}, and never with the full available dataset. Previous efforts have relied on evolutionary models of brown dwarfs and/or grid-based searches \citep{GJ229Disc}.

The spectrum of GJ 229B shows some unusual features, particularly weak H$_2$O and CH$_4$ features in the Y and K-bands that are more consistent with an earlier spectral type than the T7 type indicated by the J and H-band peaks in the spectrum \citep{Burgasser06}. This may be due to an unusually low metallicity, but other studies suggest that the GJ 229 system has near-solar metallicity \citep{GJ229Prop}. GJ 229B also shows a higher than expected abundance of CO \citep{GJ229Keck}, and measurements of its effective temperature have varied significantly, from 700-1000 K.

Recently, several studies have examined the dynamics of the GJ 229 system in order to obtain a better mass and orbital fit \citep{GJ229Brandt,GJ229Feng}. However, these results are almost completely inconsistent with each other, and neither is strongly consistent with the original imaging-based estimates given by \cite{GJ229Disc}. In light of this disagreement, an independent, spectrum-based analysis would be a valuable addition to try to resolve this tension.

GJ 229B is also notable among spatially-resolved objects for its small separation from its host star, which is on the order of tens of AU compared with hundreds or thousands of AU for many brown dwarf companions. This places it in the overlap regime between the expected stellar companion distribution and the expected planetary companion distribution \citep{RM13}. Planet formation processes at this distance are expected to yield an elevated C/O ratio in the atmosphere \citep{ObergCO}, so a measurement of this ratio will provide clues to the formation process and location of this object.

In this paper, we present a set of atmospheric retrievals of GJ 229B with the APOLLO spectrum retrieval code using the full available spectroscopic dataset. In order to explore possible limitations of the model, we perform the retrieval using two different prescriptions of the temperature-pressure profile: a five-parameter analytic model and a fifteen-layer ``free'' retrieval. To explore potential degeneracies, we also compare completely unconstrained retrievals with retrievals that alternately keep the radius and surface gravity fixed. These results enable us to place improved constraints on the mass, effective temperature, and age of GJ 229B, as well as providing fits to molecular abundances, in particular the C/O ratio.

This paper is organized as follows: the current published observations and estimated physical and orbital properties of GJ 229B are discussed in Section \ref{observe}, the APOLLO retrieval code is described in Section \ref{apollo}. The specific methods we apply to GJ 229B are described in Section \ref{sec:methods}, with our retrieval results in Section \ref{sec:results}. We discuss the implications of these results in Section \ref{sec:discuss} and summarize our conclusions in Section \ref{sec:conclude}.

\section{Observations and Previous Analyses of GJ 229B}
\label{observe}





The discovery of GJ 229B was first reported by \cite{GJ229Disc}. It was identified as a brown dwarf companion based on its integrated spectral energy distribution (SED) and common proper motion with GJ 229A, which placed its luminosity below that of low-mass stars. They found a best-fit effective temperature of 700 K from the SED and provided a mass estimate of 20-50 $M_J$ based on cooling models. The companion paper \cite{Oppenheimer95} presented the first infrared spectrum of GJ 229B based on observations from the Palomar Observatory, covering the YJHK range of 1.0-2.5 ${\rm \mu m}$. This spectrum showed significant methane absorption indicating an effective temperature of $<$1000 K.

The exact spectral type of GJ 229B is somewhat ambiguous. The initial definition of the T spectral class \citep{Burgasser02} fit GJ 229B as a normal type T6.5V dwarf. Some catalogs listed it as early as T6 \citep{Leggett02}, but later observations indicated unusual feature strengths, and it was reclassifed to T7pec in the revised classification of \cite{Burgasser06}. These ambiguous features include strong CH$_4$ features at 1.3 and 1.6 ${\rm \mu m}$, indicating the T7 type, but unusually weak CH$_4$ and H$_2$O features at 1.1, 1.4, 1.9, and 2.2 ${\rm \mu m}$, which are more typical of spectral types T5-T6. 

The peculiar spectrum was attributed to an unusually low metallicity of -0.3 to -0.5 dex by \citet{GY00}. At least one measurement of the metallicity of the host star GJ 229A is consistent with such a low value \citep{Leggett229A}; however, a more recent measurement by \cite{GJ229Prop} finds a higher metallicity for GJ 229A based on a measurement of $[{\rm C/H}]=+0.13\pm0.07$. Similarly, they use a metallicity of $[{\rm M/H}]=+0.13$ for their best-fit model with GJ 229B, leaving the correct value highly uncertain.

\subsection{Estimated Properties}

\begin{table}[htb]
    \centering
    \begin{tabular}{l | r | l | c}
    \hline
    Property       & Estimate      & Method          & Source \\
    \hline
    Radius ($R_J$) & 0.85-1.33     & Grid retrieval  & 1 \\
    Age (Gyr)      & 0.3-3.0       & Host star properties & 1 \\
    \hline
    ${\rm log}\,g$ & 5.3$\pm$0.2   & Grid fit        & 2 \\ 
                   & 4.9-5.35      & Grid fit        & 3 \\
                   & 4.5-5.5       & Evolution       & 4 \\
                   & 3.5           & Grid fit        & 5 \\
                   & 4.75-5.0      & Grid fit        & 1 \\
    \hline
    Separation (AU) & 44           & Imaging (projected) & 6 \\
                   & 34.7          & Astrometry      & 7 \\
                   & 19.4          & Radial velocity & 8 \\
    \hline
    Mass ($M_J$)   & 20-50         & Cooling rates   & 6 \\
                   & 40-55         & Cooling rates   & 2 \\
                   & 30-55         & Grid fit        & 3 \\
                   & 17-40         & Grid fit        & 1 \\
                   & $70\pm5$      & Astrometry      & 7 \\
    $M{\rm sin}\,i$ (M$_J$) & $1.62^{+0.66}_{-0.18}$ & Radial velocity & 8 \\
    \hline
    $T_{\rm eff}$ (K) & 700        & SED             & 6 \\
                   & $<$1000       & Chemistry       & 9 \\
                   & 1000$\pm$200  & Photometry      & 10 \\
                   & $<$1000       & Grid fit        & 2 \\
                   & 960$\pm$70    & Grid fit        & 3 \\
                   & 870-1030      & Evolution       & 4 \\
                   & 1000          & Grid fit        & 5 \\
                   & 750-900       & Grid fit        & 1 \\
    \hline
    \end{tabular}
    \caption{Published properties of GJ 229B. Sources: (1) \cite{GJ229Prop}, (2) \cite{Allard96}, (3) \cite{Marley96}, (4) \cite{Saumon00}, (5) \cite{Leggett229A}, (6) \cite{GJ229Disc} (discovery paper), (7) \cite{GJ229Brandt}, (8) \cite{GJ229Feng}, (9) \cite{Oppenheimer95}, (10) \cite{Tsuji96}.}
    \label{tab:lit}
\end{table}

Several early efforts at understanding GJ 229B shortly after its discovery focused on qualitative or grid-based fits using synthetic spectra from generic brown dwarf models. Some of these efforts also made their own estimates of the bulk properties of the object, primarily based on cooling models, and some found significant evidence of non-equilibrium chemistry--most notably a greater than expected abundance of CO. These early spectroscopic analyses still constitute the bulk of the available literature regarding the physical properties of GJ 229B. A summary of the published results is given in Table \ref{tab:lit}.

\cite{Tsuji96} fit a synthetic spectrum to a photometric broadband SED for GJ 229B extending from 0.8 to 10 ${\rm \mu m}$ (using the same models adopted by \citealt{GJ229Disc}). Their fit also gave $T_{\rm eff}=1000$ K, albeit with a large uncertainty of $\pm$200 K. They also studied the possibility of dust formation in the atmosphere, motivated by evidence of dust formation in late M-dwarfs; however, they found that the best fit required a dust-free model. They interpreted this as a dusty atmosphere with the dust gathered into low-altitude clouds, while the opacity was dominated by volatile gases at high altitudes, especially methane.

\cite{Allard96} paired a coarse SED fit based on effective temperature and gravity together with cooling models. They found a most probable mass range of 40-55 $M_J$, and they set an upper limit on effective temperature of $T_{\rm eff}<1000$ K. However, their grid did not extend to low enough temperatures to establish an exact fit. They also found no evidence of atmospheric aerosols based on the spectrum colors, but suggested a second possible explanation for this finding consistent with the current model of rainout of cloud condensates.

\cite{Marley96} fit the observed spectrum of GJ 229B with a grid of synthetic spectra with the aid of a stellar evolution code. They found a much more precise estimate of the object's effective temperature of 960$\pm$70 K, but their mass estimate was similar to the others at 30-55 $M_J$.

\cite{Saumon00} presented a new NIR spectrum of GJ 229B from UKIRT, following up on the earlier work of \cite{GJ229UKblue}. They performed a grid fit to this spectrum incorporating specific molecular abundances, based on three specific evolutionary models. They were able to positively detect a depleted, non-equilibrium abundance of NH$_3$ in the atmosphere of GJ 229B, along with the previously observed elevated abundance for CO \citep{GJ229UKred}. This non-equilibrium chemistry was attributed to vertical mixing from deeper layers of the atmosphere. Their analysis also suggested an overall metal-poor composition.

\cite{Leggett229A} fit a grid of models that included overall metallicity, along with measuring the metallicity of GJ 229A. They found a best-fit metallicity of -0.5 dex and, again, an effective temperature of 1000 K. However they also fit a very low surface gravity of ${\rm log}(g)=3.5$, suggesting a mass possibly smaller than the deuterium-burning limit.

\subsection{Current State of the Art}

Few recent studies have been done on GJ 229B using modern techniques. The most notable spectroscopic study is that of \cite{GJ229Prop}, who performed a grid fit of surface gravity, radius, and effective temperature using the Uniform Cloudy Models, with additional constraints based on the age of the object. For this purpose, they adopted a prior estimate of the age of the system of 0.3-3.0 Gyr. The lower limit was based on the activity level of the host star, while the upper limit was based on the kinematics of the system as an apparent young-disk star. The same study reevaluated the metallicity of both components of the system to +0.13 dex.

Notably, this study found a significantly lower effective temperature than earlier ones, with acceptable fits in the allowed age range from 750 K to 900 K, the best fits being at 800-850 K. All of the good-fit models had ${\rm log}(g)=4.75$-5.0, consistent with the mean of the previous estimates. Unlike many previous studies, which assumed a brown dwarf radius of $\sim$1 R$_J$, they also made an independent retrieval of the radius, using it as the fitting parameter to find the best fit to the spectrum for each temperature-gravity pair. The resulting range of radii was 0.87-1.33 $R_J$, corresponding to masses of 17-40 $M_J$.

The other recent studies of GJ 229B have focused on new dynamical measurements of its mass. GJ 229B was originally reported to have a projected separation of 44 AU from GJ 229A \citep{GJ229Disc}, large enough that a long baseline would be needed to make an accurate determination of its orbit. With 25 years now passed since its discovery, such a long baseline is available. However, these studies have proved particularly troubling in that they show large disparities in their estimates of GJ 229B's mass and, unusually, its separation from its host star.

\cite{GJ229Feng} performed a radial velocity analysis to search for a long-term trend that could be fit to the orbit of GJ 229B. They reported an RV fit of $a=19.4$ AU and a low eccentricity $e=0.07$. This is troubling because it would be impossible for such an orbit to produce the original observed projected separation of 44 AU. They also reported a very low value of $M{\rm sin}\,i$ of $1.62^{+0.66}_{-0.18}$ M$_J$, which they interpreted as a more massive object with a nearly face-on orbital geometry. While this is possible, it would require an improbably low inclination of $i\lesssim5^\circ$ to be consistent with a brown dwarf mass.

Meanwhile, \cite{GJ229Brandt} reported an astrometric fit of GJ 229B's orbit based on \textit{Hipparcos} and \textit{Gaia} data. They reported an orbital distance of $a = 34.7$ AU, with a much larger eccentricity of $e=0.846$ and a less stringent limit on inclination of $i<23^\circ$. However, they also found a very high mass of $70\pm5\,M_J$. This is inconsistent with evolutionary models (see Section \ref{sec:evo}), although they suggested that it could be consistent if GJ 229A is older than previously thought.

These two orbital fits are dramatically inconsistent with each other. The astrometric fit is consistent with the original observations, but includes other improbable features, and the RV fit is completely inconsistent with the original observations. This may be a result of the length of the available baseline; the reported radial velocity period is $126\pm26$ years, while the reported astrometric period is more than twice as long at $263_{-21}^{+29}$ years. Both of these are long enough that it may simply be that we still do not have a long enough baseline for an accurate orbital fit. Notably for the astrometric fit, the observed position angle of GJ 229B changed by only $7^\circ$ from 1996 to 2011.

Further complicating this picture is the discovery of two planets orbiting GJ 229A at distances of $<1$ AU \citep{GJ229Planet,GJ229Feng}. These planets have reported $M {\rm sin}\,i$ of 8 and 10 $M_\oplus$, suggesting Neptune-type planets; but if they have similarly face-on orbits, they could be as massive as Jupiter. With a complex and potentially quite massive planetary system, a revised radial velocity fit for the system could resolve the discrepancy between the two results, possibly retrieving an orbit consistent with the astrometric fit. Alternatively, another interpretation is that the small orbit indicated by the RV fit is evidence of an as-yet undiscovered third planet of a few Jupiter masses orbiting GJ 229A, while the orbit of GJ 229B, being 2$-$3 times longer, would not be readily detectable as an RV trend.

In any case, further observation is needed to accurately resolve the dynamical mass and orbit of GJ 229B. However, atmospheric retrievals combined with evolutionary models can provide some clues for solving this puzzle.

\section{The APOLLO Spectroscopic Retrieval Code}
\label{apollo}

APOLLO is an open source radiative transfer and spectroscopic retrieval code first described in early form by \cite{ApolloIntro}.\footnote{Source code available at https://github.com/alexrhowe/APOLLO. This work uses APOLLO Version 0.12.1.} In general, APOLLO builds forward models by modeling the radiative transfer of light through a planetary (or brown dwarf) atmosphere. For emission spectra (as are used in this work), this is a plane-parallel absorbing and emitting atmosphere. APOLLO can also calculate transit spectra, modeling the passage of starlight through the atmosphere, taking the geometry of the atmosphere into account. In both cases, Rayleigh scattering is included as a source of absorption. The retrieval is performed using a Markov-Chain Monte Carlo (MCMC) algorithm provided by the Emcee Python package \citep{Emcee}.

\subsection{The Forward Model}

The radiative transfer is simulated by dividing the atmosphere into many discrete layers (70-100 by default, depending on the configuration), with the temperature-pressure profile computed separately and interpolated to match these layers. For each layer, the atmospheric emission and optical depth are computed for each wavelength point. The contributions of each layer are summed up, and in the case of an emission spectrum, they are averaged over the outward-facing hemisphere by an 8-point Gaussian quadrature method. (For a transit spectrum, the rays are assumed to be parallel.)

Notably, APOLLO offers the choice of two radiative transfer functions: a simple one-stream approximation that includes only outward flux and a hemispherical mean two-stream approximation \citep{Toon} that includes a more careful treatment of multiple scattering. For GJ 229B, with the observations being mostly in the infrared and a cloudless atmosphere model, we find negligible differences between the results of these two algorithms, although we use the two-stream approximation by default.

The radiative properties of the model atmosphere are determined by the molecular absorption and scattering cross sections, the temperature-pressure profile of the atmosphere, and any prescription for clouds that is included. The molecular cross sections are read from a precomputed grid and may be optionally pre-binned to a lower resolution to reduce computation time. Alternatively, the observations may be either convolved or binned to a lower resolution to better accommodate low-resolution tables; however, neither was necessary for this study. 

For this work, we used the cross section tables of \cite{Freedman} compiled based on the HITRAN and ExoMOL line lists. The tables were interpolated to a grid of 18 evenly-spaced log-pressure points from 300 to 10$^{-6}$ bars and 36 evenly-spaced log-temperature points from 75-4000 K, a structure chosen as a regularization of the original grid at which the tables were calculated to minimize interpolation. We used a default spectral resolution of R=10,000 for the cross-section tables. These working tables were then interpolated to the temperature and pressure of each layer in the model atmosphere to compute the forward model spectrum. For the purposes of calculating the effective temperature, a separate set of tables are used at much lower resolution (see Section \ref{sec:stats}). By default we use the same set of tables binned down to R=200 for this purpose.

The molecular composition of the atmosphere is assumed to be constant with altitude, which reduces the number of parameters needed to model the atmosphere; this is also consistent with a large fraction of the outgoing emission originating from a relatively thin layer of the atmosphere near the photosphere. We perform a free retrieval of the log-molecular abundances with a hydrogen-helium ``filler'' gas added to bring the sum of the abundances to 1. This minimizes any assumptions that need to be made about non-equilibrium chemistry in the atmosphere. For this work, we selected H$_2$O, CH$_4$, CO, CO$_2$, NH$_3$, H$_2$S, and the alkali metals Na+K.

For the temperature-pressure profile of the atmosphere, APOLLO again offers the choice of two prescriptions. One is the five-parameter radiative equilibrium model of \cite{Guillot}, while the second is a ``free'' retrieval using a layer-by-layer temperature-pressure profile with specified maximum and minimum pressure levels and a specified number of evenly-spaced log-pressure points (as well as an inter-layer smoothing parameter, described below). For this work, we used a 15-layer profile running from 300 to 0.001 bars, assuming the atmosphere to be isothermal at higher altitudes. For GJ 229B, we find that the layer-by-layer profile produces a significantly better fit to the observed spectrum, which appears to be related to increased degrees of freedom for specific pressure levels.

Finally, we do not include clouds in our forward models of GJ 229B. Even allowing for large uncertainties, the effective temperature is well below that of the L/T transition where clouds are expected to become a major contributor to brown dwarf spectra. This also helps us by reducing the number of free parameters in the retrieval. Nonetheless, APOLLO includes several options for different cloud/aerosol prescriptions in the forward models. In the version used for this work (0.12.1), the options are (1) no clouds, (2) an opaque cloud deck cutting off all transmission below a certain pressure level, (3) a Gaussian cloud model with a fixed particle size and particle abundance following a Gaussian profile in log-pressure space, and (4) a slab cloud model with a fixed particle size and a constant particle abundance between two pressure levels. In the latter two cases, the aerosol extinction is computed using Mie theory. Additional cloud models are in development for future versions of the code \citep{HD106}. We summarize the configuration settings we using in this study in Table \ref{tab:config}. 

All of the model parameters in APOLLO may also be fixed to a particular value and removed from the retrieval calculation by setting the standard deviation on the prior to zero (regardless of the type of prior).

\begin{table}[htb]
    \centering
    \begin{tabular}{l | l | l}
    \hline
    Setting & APOLLO Options & This Work \\
    \hline
    Table resolution        & Free & $R=10,000$ \\
    SED Table resolution & Free & $R=200$ \\
    Observation convolution & Free & None \\
    Observation binning     & Free & None \\
    Walkers                 & $>2\cdot N_{\rm params}$ & $8\cdot N_{\rm params}$ \\
    Steps                   & Free & 30,000 \\
    \hline
    Prior                   & Uniform & Uniform \\
                            & Gaussian & \\
    \hline
    Radiative transfer & One-stream & Testing \\
                       & Two-stream & Yes     \\
    \hline
    Filler gas      & Available tables & H$_2$+He \\
    Retrieved gases & Available tables & H$_2$O, CH$_4$, CO, CO$_2$, \\
                    &                  &  NH$_3$, H$_2$S, Na+K \\
    \hline
    T-P profile & 5-parameter    & Yes      \\
                & N-Layer        & 15 layers, smoothed, \\
                &                & 300$-$0.001 bars \\
    \hline
    Aerosols    & None              & None \\
                & Opaque cloud deck &      \\
                & Gaussian, fixed particle size & \\
                & Slab, fixed particle size     & \\
    \hline
    \end{tabular}
    \caption{Configuration settings for APOLLO used in this work.}
    \label{tab:config}
\end{table}

\subsection{Statistical Methods}
\label{sec:stats}

Along with the forward model parameters, we include several ``statistical'' parameters in the retrieved parameter set to account for potential errors or uncertainties in the observations. These ensure that the forward models align correctly with the observations and are accurately reporting the subsequent goodness-of-fit. The wavelength calibration parameter, $\Delta L$, shifts the spectrum to higher or lower wavelengths to account for errors in the wavelength calibration of each observation. The error bar scaling parameter, ${\rm ln}(f)$, applies a constant multiplier to the reported uncertainties, which corrects for unaccounted-for sources of noise or uncertainty in the spectrum. This is especially important for GJ 229B, where the original observational uncertainties are unavailable, and we had to estimate them from the high-frequency variance of the spectra (see Section \ref{sec:data}). For retrieval on spectra containing multiple data sets, we include one $\Delta L$ and one ${\rm ln}(f)$ parameter for each data set.

Band-to-band calibration may also be a source of uncertainty between datasets.  If a spectrum contains multiple datasets, as is the case with GJ 229B, APOLLO includes parameters to scale them relative to each other, labeled $S_n$. For the full GJ 229B dataset of four spectra, we include three such parameters with the YJHK spectrum as the reference point; uncertainty in the absolute flux calibration of this spectrum is also incorporated into the retrieved uncertainty on the radius parameter.

A final source of statistical uncertainty comes from the forward model itself, specifically, the temperature-pressure profile. A completely free layer-by-layer T-P profile is prone to overfitting, leading to an unphysical profile with many large swings in temperature between layers. To suppress this behavior, APOLLO includes an optional smoothing hyperparameter, $\gamma$, as used by \cite{Line2015}. This parameter effectively penalizes the model prior for large second derivatives in the T-P profile according to the equation,

\begin{equation}
    {\rm ln}({\bf T}) = -\frac{1}{2\gamma}\sum_{i=1}^{N}\left(T_{i+1} - 2T_i + T_{i-1}\right)^2 - \frac{1}{2}{\rm ln}(2\pi\gamma).
\end{equation}

Spectroscopic retrievals also require a prior distribution of the expected atmospheric properties. APOLLO includes two prescriptions for priors on most model parameters: uniform priors and Gaussian priors, in both cases with fixed minimum and maximum values set by the user. The exception to this rule is the T-P profile smoothing parameter $\gamma$, which uses an inverse gamma distribution for a prior with parameters set to $\alpha=1$ and $\beta=5\times10^{-5}$. In this work, we use uniform (or log-uniform) priors for all of the other parameters.

In addition to the priors on retrieved parameters, APOLLO includes two internal priors on the surface gravity and the derived mass. These are not included in the prior probability calculation except to return a zero prior probability if they are outside the allowed range$-$effectively a uniform prior. Since mass is fully determined by radius and surface gravity, we do not use it to add to a non-zero prior. For the mass, in this work, we limit it to the brown dwarf range of 13-80 M$_J$. Finally, because an atmosphere that is too thick and low-density may fail to converge to a finite radius, there is an additional derived limit on the surface gravity such at the scale height of the atmosphere may not be more than 5\% of the planet's ``core'' radius. This is suitable for most known exoplanets and brown dwarfs -- although it may fail for the very low-density ``super-puffs,'' which have anomalously low densities of $<0.1\,{\rm g\,cm^{-3}}$ \citep{SuperPuff}, no plausible planet of Saturn-mass or greater would have such a large scale height.

APOLLO fits the forward model to the input spectrum using a Markov-Chain Monte Carlo (MCMC) method provided by the Emcee Python package. The MCMC approach is a Bayesian method that randomly samples the parameter space with an ensemble of ``walkers'' or ``chains'' that traverse the space iteratively, the probability distribution for each walker's next step being weighted by its relative goodness of fit on the current step. The goodness of fit is computed according to a Bayesian likelihood function based on a modified chi-squared statistic weighted by the error bar correction term:

\begin{equation}
    {\rm ln}\,L({\bf f_{\rm obs}}|{\bf x}) = -\frac{1}{2}\sum_{i=1}^n\frac{\left(f_{{\rm obs},i}-f_{{\rm mod},i}({\bf x})\right)^2}{s_i^2} - \frac{1}{2}{\rm ln}(2\pi s_i^2),
\end{equation}

where ${\bf x}$ is the vector of model parameters, ${\bf f}_{\rm obs}$ are the observed fluxes, ${\bf f}_{\rm mod}({\bf x})$ are the model fluxes, and $s_i$ are the adjusted uncertainties with the correction parameter ${\rm ln}\,f$ included:

\begin{equation}
    s_i^2 = \sigma_i^2 + f.
\end{equation}

Note that with this correction parameter, it is possible for the log-likelihood to be greater than zero. With sufficiently many walkers and steps, the weighted random sample should approach the best fit to the observations, with the same distribution as the actual likelihood of the parameters. In this work, we use a standard prescription of 8 times as many walkers as model parameters \citep{Emcee}, and we allow the walkers to run for 30,000 steps, which we verified as sufficient by plotting their trajectories to show that it leads to a stable posterior distribution. The posterior statistics are computed from the last 10\% of samples.

Our model parameters and priors for GJ 229B are listed in Table \ref{tab:model}, including the two derived priors. We set the limits broadly as often as possible, using only the limits necessary to prevent the atmosphere model from straying into unphysical territory. However, we did use somewhat narrower bounds on the mass and radius of the object based on the grid fit of \cite{GJ229Prop}.

APOLLO reports four derived parameters along with the retrieved model parameters. These are the mass, the metallicity of the atmosphere, the C/O ratio, and the effective temperature. The first three can be computed from the model parameters; metallicity and C/O ratio are computed by summing the associated molecular abundances, and mass can be found directly from radius and surface gravity. However, we calculate the effective temperature differently, using the per-unit-area SED from the object's surface, thus bypassing potential errors introduced by distance and flux calibration. To do this, APOLLO computes a lower-resolution model spectrum ($R=200$ by default) over a much wider wavelength range of 0.6$-$30 ${\rm\mu m}$. It is assumed that nearly all of the flux from typical planets and brown dwarfs will fall in this range, so $T_{\rm eff}$ can then be computed with the Stefan-Boltzmann law.

\begin{table}[htb]
    \centering
    \begin{tabular}{l | l}
    \hline
    Parameter & Prior Range \\
    \hline
    Model Parameters & \\
    Radius ($R_J$) & $0.85-1.33$ \\
    log($g$) & $4.50-5.25$ \\
    log($X_i$) & $-12.0--2.0$ \\
    \hline
    Parametric T-P Profile & \\
    $T_{\rm int}$ (K) & $75-4000$ \\
    log($\kappa_{\rm IR}$) & $-5.0-+5.0$ \\
    $\gamma_1$ & $0.01-2.0$ \\
    $\gamma_2$ & $0.01-2.0$ \\
    $\alpha$ & $0.0-1.0$ \\
    \hline
    Layer-by-Layer T-P Profile & \\
    $T_1-T_2$ (K) & $100-4000$ \\
    $T_3-T_6$ (K) & $100-2000$ \\
    $T_7-T_9$ (K) & $100-1000$ \\
    $T_{10}-T_{15}$ (K) & $100-750^a$ \\
    $\gamma={\rm InvGamma}(\alpha=1,\,\beta=5\times10^{-5})$ & $0-1000$ \\
    \hline
    Statistical Parameters & \\
    $S_0,S_2,S_3$ & $0.9-1.1$ \\
    $\Delta L$ (nm) & $-10.0-+10.0$ \\
    ln($f$) & $0.01\cdot{\rm min}(\sigma)-100\cdot{\rm max}(\sigma)$ \\
    \hline
    Derived Parameters & \\
    Mass ($M_J$) & $13-80$ \\
    Scale height & $0.00\cdot R-0.05\cdot R$ \\
    ${\rm [M/H]}$ (dex) & None \\
    C/O ratio & None \\
    $T_{\rm eff}$ (K) & None \\
    \hline
    \end{tabular}
    \caption{Forward model parameters for GJ 229B. $^a$The relatively low temperature limits in the upper atmosphere were used to aid convergence in difficult cases. They were tested by comparison with retrievals with broader limits.}
    \label{tab:model}
\end{table}

\pagebreak

\section{Methods}
\label{sec:methods}

\subsection{Data Set}
\label{sec:data}

We use four data sets at different wavelengths in our retrieval of GJ 229B, falling in a range from 0.524 ${\rm \mu m}$ to 5.093 ${\rm\mu m}$. We initially analyze the combined data for all of the observations, and then examine the impact of only using subsets of the full data set.

GJ 229B has been observed several times in the near infrared. For our retrievals we use the spectrum taken by \cite{GJ229UKblue}, since this spectrum is available for download from the L and T Dwarf Data Archive by S. Leggett\footnote{http://staff.gemini.edu/$\sim$sleggett/LTdata.html}, along with the other data sets we use. This spectrum was taken by the CGS4 spectrograph on the UKIRT telescope and fully covers the range of 1.0245 to 2.5190 ${\rm \mu m}$. Thus, it includes much of the Y band (including the Y band spectral peak characteristic of T-dwarfs) along with the J, H, and K bands.

A great many brown dwarf spectra are observed in or near the 1.0-2.5 ${\rm \mu m}$ wavelength range (e.g. \citealt{SloanBDs}), and the observations of \cite{GJ229UKblue} in roughly the same wavelength range are the largest of our four data sets. Therefore, we make this data set the core of our analysis of GJ 229B. We performed one retrieval, specifically, a free retrieval with a layered T-P profile, on this set only for comparison to determine how well APOLLO will be able to constrain the properties of other brown dwarfs and giant planets. We also performed a retrieval on this NIR spectrum combined with each of the other data sets individually to help determine what wavelengths are most useful for future observations. The remainder of our retrievals with different parameterizations as discussed below were performed on all four data sets together.

Joined directly to the blue end of the YJHK spectrum is the space-based spectrum taken by the STIS spectrograph on the \textit{Hubble Space Telescope} \citep{GJ229HST}, covering the remainder of the Y band along with the Z, I, and R bands, blueward to 0.524 ${\rm \mu m}$. Flux in this region is much lower due to the low effective temperature of GJ 229B, but it is of particular interest because it covers strong alkali metal lines that can be used to constrain the surface gravity of the object.

At longer wavelengths, we include the L band spectrum taken by \cite{GJ229Keck} using the Near Infrared Camera on the Keck I telescope, covering the range if 2.98 to 4.14 ${\rm \mu m}$, and a second observation in the M band from UKIRT \citep{GJ229UKred} from 4.51 to 5.09 ${\rm \mu m}$.

Of these four data sets, only the M band spectrum has published uncertainties, which are not available in numerical form. To estimate the uncertainties in the observations we employ the same strategy used by \cite{GJ229UKred} of taking the RMS deviations of the observed spectrum from a smoothed version of the same spectrum. We smoothed the spectra by convolving each one with a Gaussian kernel with a full width at half-maximum of 10 times the published resolution. The residuals between the original and smoothed spectra were then convolved with the same kernel to produce our adopted uncertainties. The resulting uncertainties were similar to those of \cite{GJ229UKred} in the M band. This fiducial estimate of uncertainty combined with the error bar scaling parameters in the retrieval should allow us to adequately account for the unknown uncertainties in the original observations.

Another complication of these data is that the absolute calibrations of the spectra and their uncertainties are not known. While we retrieve for the ``band-to-band'' differences in calibration between data sets, the retrieval cannot account for the uncertainty in the absolute flux reaching the instrument, nor can it account for the uncertainty in the distance to the object. Because these factors do not change the shape of the spectrum (unlike the temperature and surface gravity), they are both degenerate strictly with the radius parameter.

To account for this, we re-scale our retrieved radii and their uncertainties based on an independent photometric flux measurement. For this purpose, we choose the MKO J-band magnitude for GJ 229B derived from the 2MASS J-band observations \citep{Faherty}. We re-scale the radius parameter (and thus the mass) such that the forward model spectrum has J-magnitude equal to the photometric measurement. We then propagate the uncertainties in flux and distance into the retrieved radius and mass uncertainties.

\subsection{Model Selection}

One of the greatest challenges for retrievals of brown dwarfs and giant planets comes from the degeneracies observed between radius, surface gravity, effective temperature, and metallicity (e.g. \citealt{Burrows01}). This is further complicated by the fact that we do not retrieve on $T_{\rm eff}$ directly, but on the radius and the temperature-pressure profile, introducing an additional degeneracy.

In order to attempt to disentangle the effects of these degeneracies, we ran retrievals on our full spectrum of all four data set with both the layered and parametric temperature-pressure profiles, with several retrievals for each profile$-$one free retrieval and three with certain parameters fixed: radius alone, surface gravity alone, and both radius and surface gravity. For the fixed parameters we selected a radius of 1 $R_J$ and a surface gravity of ${\rm log}\,g = 4.87$, corresponding to a mass of approximately 30 $M_J$, in the middle of the published estimates. This model set allows all combinations of the relevant parameters to be fixed or varied except for the metallicity, thus testing the effect of potential degeneracies among them.

\subsection{Goodness of Fit Statistics}

To measure the quality of our retrieved spectra, we examine three measures of goodness of fit. The log-likelihood is the parameter used by APOLLO to determine the best fit, that is, the Bayesian likelihood of the sample incorporating the prior. We also compute the reduced chi-squared statistic for each model. And finally, for a more intuitive picture of the goodness of fit, we compute the root-mean-square error of the forward model spectrum from the observations, measured in standard deviations. Because these goodness-of-fit statistics are based on the uncertainties in the spectrum, and we include a scaling of those uncertainties as one of the parameters, we compute these statistics both with and without the error bar scaling parameters included (except for the log-likelihood, for which it is the same).

\subsection{Evolutionary Model Comparison}
\label{sec:evo}

A further source of guidance regarding the bulk properties of GJ 229B are evolutionary models of brown dwarfs. These are limited by the large uncertainty in the age of the system, but they can still indicate the expected theoretical ranges of mass, radius, gravity, and effective temperature for the object. As discussed in Section \ref{observe}, several past estimates of GJ 229B's properties have been made from evolutionary models.



The consistency of our retrieved parameters with evolutionary model predictions can provide an additional indicator of the accuracy of our inferred atmospheric models along with the goodness of fit. If a retrieval is a good fit to evolutionary models, this is evidence that it is more likely to be a reliable parameterization. Similarly, the retrievals on the partial data sets can be compared with evolutionary models to predict which wavelength ranges will be most diagnostic of brown dwarf and planetary bulk properties such as surface gravity. These estimates also provide an additional check for the dynamical mass estimates for GJ 229B based on estimated cooling rates. We perform a fit for mass and age using the recently published Sonora-Bobcat evolutionary models \citep{Sonora} based on our retrieved radius, gravity, and effective temperature. This can also provide a better estimate of the age of the system than the stellar properties.


The Sonora-Bobcat evolutionary tables are provided at three metallicities of -0.5 dex, solar, and +0.5 dex. Based on our retrieved metallicity of [M/H]$\,=-0.07\pm0.03$, we use the solar-metallicity table for our fits. The evolutionary tables provided also assume a solar C/O ratio of 0.5, lower than our retrieved value of $1.13\pm0.03$. However, the solar value should still be correct for the interior structure of brown dwarfs, and the ratio of carbon to oxygen will be a secondary effect to metallicity on the effective temperate of the object, so the Sonora-Bobcat models should nonetheless provide the best available fit.






\section{Results}
\label{sec:results}

\subsection{Best Fit Parameters}

For each of our 12 selected models, we retrieve both atmospheric properties and certain ``bulk properties'' of GJ 229B. These include directly retrieving the radius and surface gravity, plus the derived parameters of mass and effective temperature. (The bulk properties of metallicity and C/O ratio are discussed in Section \ref{sec:chem} along with the atmospheric chemistry.) The retrievals also return posterior distributions for all of these parameters.

The goodness-of-fit statistics for our models are listed in Table \ref{tab:stats}. For clarity, we present the log-likelihood statistics in reduced form, dividing by the number of spectral points. All three of these statistics track together well across the three groups of retrievals (layered, parametric, and limited data) in the ``absolute'' measures without the error bar scaling parameters included. The lowest reduced chi-squared values correspond to the lowest RMS error values and the highest likelihoods. This suggests that our selection of the best fit model is a robust one. 

The error bar scaling parameters naturally settle into an equilibrium in which the error bars are inflated just enough to find a good fit to the data. As such, we would expect the goodness-of-fit statistics incorporating these parameters to be universally low. This is indeed what we see with $\chi_\nu^2\sim1.5-2.4$, which suggests that APOLLO is reliably finding the maximum likelihood.

\begin{table}[htb]
    \centering
    \begin{tabular}{l | l | r | r | r | r | r}
    \hline
    T-P Profile & Retrieval & Reduced Log-Likelihood & $\chi_\nu^2$ & RMS error ($\sigma$) & $\chi_\nu^2$ & RMS error ($\sigma$) \\
    \hline
    \  & \ & \ & \multicolumn{2}{c|}{Original error bars} & \multicolumn{2}{c}{Scaled error bars} \\
    \hline
    Layered     & Free              & 18.47 & 8.147 & 2.404 & 2.206 & 1.506 \\
                & Fixed $R$         & 18.46 & 10.063 & 2.509 & 2.126 & 1.543 \\
                & Fixed $g$         & 18.47 & 8.345 & 2.415 & 2.206 & 1.513 \\
                & Fixed $R$ and $g$ & 18.45 & 9.131 & 2.532 & 2.274 & 1.534 \\
    \hline
    Parametric  & Free              & 18.19 & 17.340 & 2.952 & 1.804 & 1.301 \\
                & Fixed $R$         & 18.19 & 17.262 & 2.947 & 1.697 & 1.286 \\
                & Fixed $g$         & 18.18 & 18.665 & 2.935 & 1.477 & 1.279 \\
                & Fixed $R$ and $g$ & 18.18 & 17.325 & 2.892 & 1.578 & 1.261 \\
    \hline
    Limited Data & YJKH only        & 17.92 & 8.412 & 2.197 & 1.701 & 1.310 \\
                & IZ + YJKH$^a$     & 18.08 & 9.087 & 2.277 & 1.882 & 1.343 \\
                & YJKH + L          & 17.98 & 10.232 & 2.256 & 1.708 & 1.333 \\
                & YJKH + M          & 17.95 & 10.283 & 2.728 & 2.420 & 1.669 \\
    \hline
    \end{tabular}
    \caption{Goodness of fit for our retrievals of GJ 229B. $^a$This retrieval discarded the wavelength range of 0.52-0.80 ${\rm \mu m}$ and used narrower priors on the T-P profile. (See Section \ref{sec:riz}).}
    \label{tab:stats}
\end{table}

\begin{figure}[h!]
    \centering
    \includegraphics[width=0.99\textwidth]{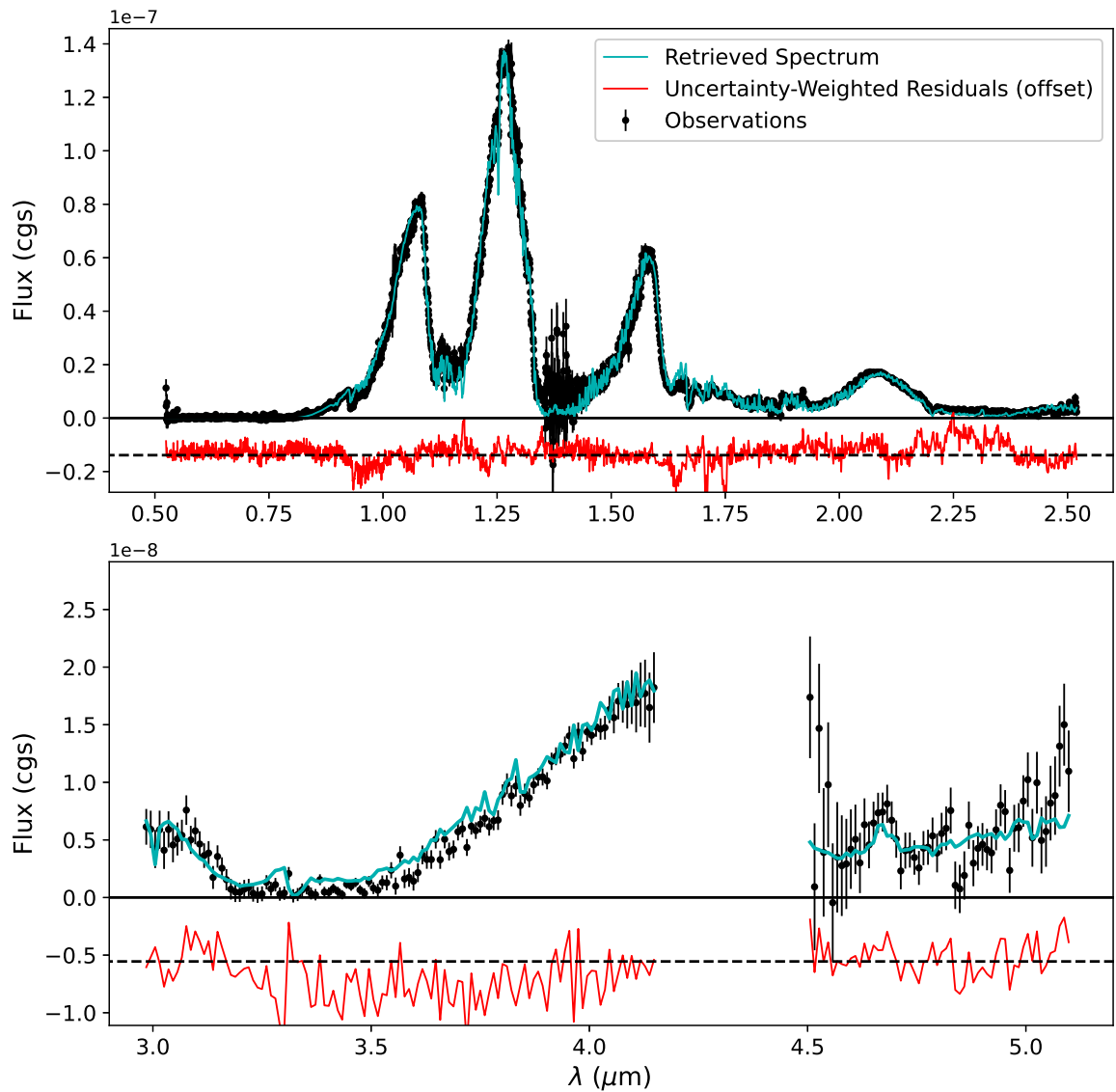}
    \caption{Best fit spectrum for the full dataset for GJ 229B. The forward model spectrum is plotted over the observational data with our fiducial error bars based on the point-to-point variance. We also plot weighted residuals offset below the spectrum. The residuals are weighted by the size of the uncertainty on the corresponding observations divided by the mean uncertainty.}
    \label{fig:GJ229B_spec}
\end{figure}

\begin{table}[htb]
    \centering
    \begin{tabular}{l | l}
    \hline
    Radius         & $1.105\pm0.025\,R_J$         \\
    ${\rm log}\,g$ & $4.93_{-0.03}^{+0.02}$         \\
    Mass         & $41.6\pm3.3\, M_J$              \\
    $T_{\rm eff}$  & $869_{-7}^{+5}$ K              \\
    ${\rm [M/H]}$  & $-0.074_{-0.030}^{+0.028}$     \\
    C/O            & $1.133_{-0.029}^{+0.027}$      \\
    \hline
    ${\rm [H_2O]}$ & $-3.430_{-0.016}^{+0.017}$     \\
    ${\rm [CH_4]}$ & $-3.352_{-0.017}^{+0.018}$     \\
    ${\rm [CO]}$   & $-3.805_{-0.113}^{+0.110}$     \\
    ${\rm [NH_3]}$ & $-4.504_{-0.024}^{+0.021}$     \\
    ${\rm [Na+K]}$ & $-5.942_{-0.014}^{+0.11}$      \\
    ${\rm [H_2S]}$ & $-5.6_{-4.5}^{+0.9}$           \\
    ${\rm [CO_2]}$ & $-7.9_{-2.7}^{+2.6}$           \\
    \hline
    \end{tabular}
    \caption{Adopted best fit atmospheric parameters for GJ 229B, best on a free retrieval of the full dataset with a layered T-P profile.}
    \label{tab:best}
\end{table}


The best fit we find to the observed spectrum by all three goodness-of-fit metrics is the ``default'' free retrieval on the full data set with a layered temperature-pressure profile. Moreover, the results of the retrievals with radius and gravity fixed with this profile are significantly closer to the best fit than to the other groups of retrievals. Thus, we adopt this retrieval for our accepted values of GJ 229B's atmospheric properties. The results for all of the parameters in this retrieval are listed in Table \ref{tab:best}.

We note that the values for some of the retrieved parameters are very tightly constrained. In particular, the effective temperature is constrained within 1\%. However, comparisons of nearly-identical forward models prove that this is a plausible result, with a 1\% difference in temperature corresponding to a difference of approximately 1 in the reduced chi-squared statistic. Similar results hold for the surprisingly narrow constraints on surface gravity.

The spectrum fit to the full data set of observations is shown in Figure \ref{fig:GJ229B_spec}. (In this figure, we plot our original estimated error bars based on the point-to-point variance in the spectrum.) The forward model shows a very good fit across the full range of wavelengths, with the largest errors occurring in regions with large amounts of noise and at the edges of the observing bands. To better illustrate the quality of the fit, we plot uncertainty-weighted residuals offset below the spectrum. These residuals are scaled inversely with the observational uncertainty divided by the mean uncertainty so that regions of high noise show less significant residuals, but this scaling results in some additional high-frequency artifacts in the residuals such as the apparent narrow features at $\sim$1.7 ${\rm \mu m}$.

\subsection{Parametric Temperature Profile}

\begin{figure}[h!]
    \centering
    \includegraphics[width=0.99\textwidth]{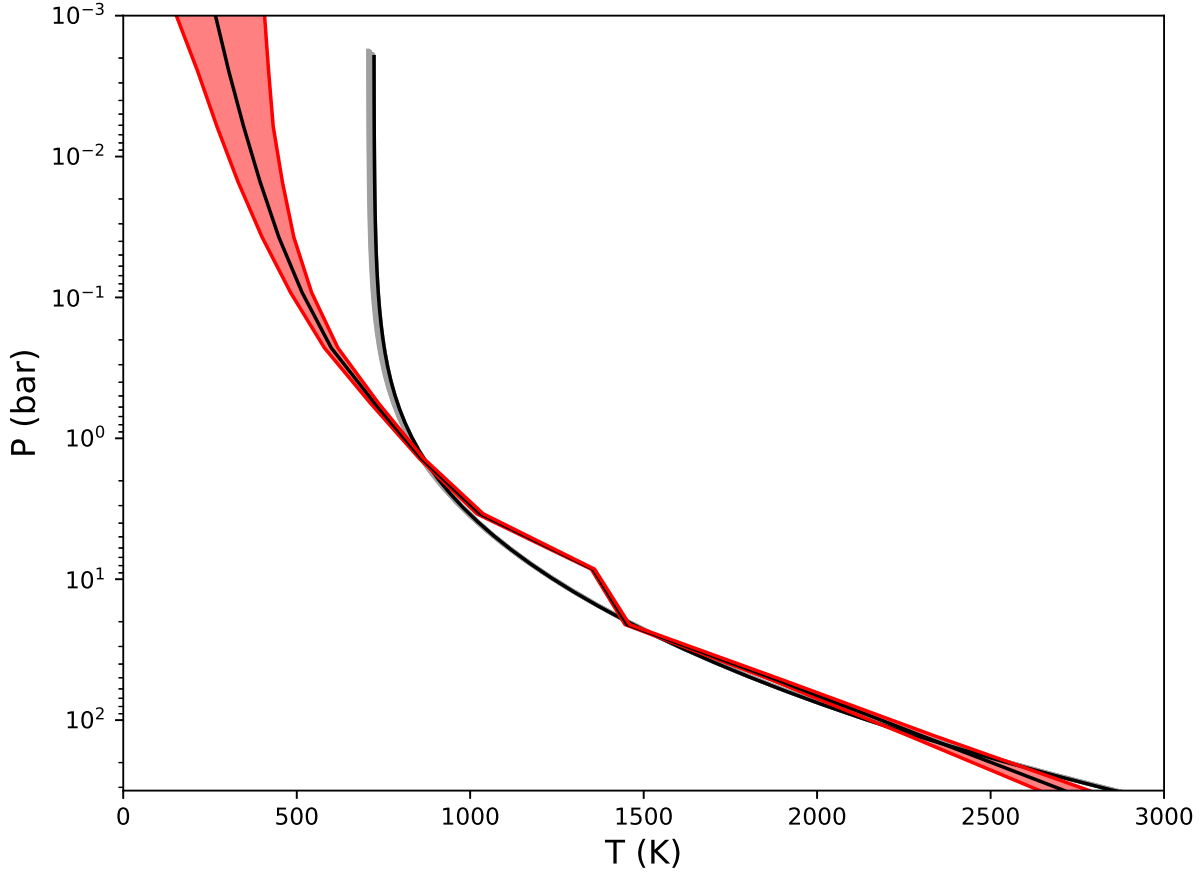}
    \caption{Retrieved temperature-pressure profiles for the full dataset for GJ 229B. Red: layered profile with 1-$\sigma$ uncertainties from our best-fit retrieval. Gray: sampling of the posteriors of the best-fit parametric T-P profile. Median profiles are shown in black.}
    \label{fig:GJ229B_TP}
\end{figure}

\begin{figure}[h!]
    \centering
    \includegraphics[width=0.99\textwidth]{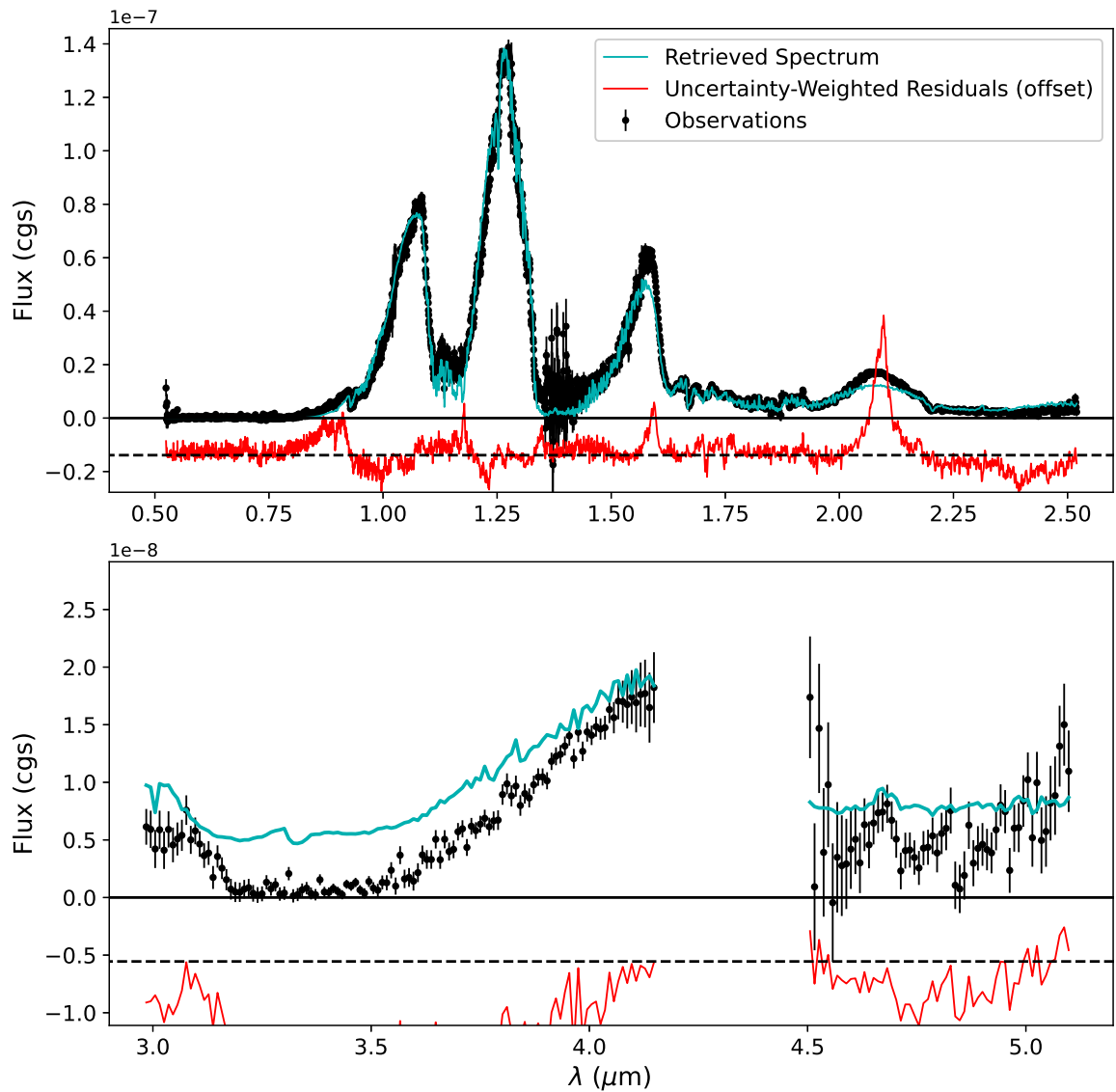}
    \caption{Best fit spectrum for GJ 229B with a parametric T-P profile. Uncertainty-weighted residuals are plotted offset below the spectrum.}
    \label{fig:paramspec}
\end{figure}

For completeness, we also show the results of the parametric T-P profile calculation for the best fit model in that group, which in this case is the retrieval with fixed radius and surface gravity, although all of them are very close to one another. The best-fit layered T-P profile (red) is plotted in Figure \ref{fig:GJ229B_TP}, along with the best fit parametric T-P profile (gray). Our best fit forward model spectrum using a parameterized T-P profile is plotted against the observations of GJ 229B in Figure \ref{fig:paramspec} (again, plotting our original uncertainties based on the point-to-point variance in the spectrum). This spectrum is a generally poorer fit at all wavelengths, with inaccurate Y, H, and K-band peak heights and a significant systematic offset across most of the L-band. 

The results for the parametric profile show significant differences in several retrieved parameters, mainly the surface gravity, effective temperature, and metallicity. As discussed in Section \ref{sec:evofits}, the surface gravities for the parametric profile (when they are allowed to vary) are significantly lower at ${\rm log}\,g\approx4.5$, compared with ${\rm log}\,g=4.93$ for the layered profile. A lower surface gravity suggests that the largest proportion of flux should originate from higher in the atmosphere due to the greater optical depth. However, this is counterbalanced by a much lower metallicity of -0.3 $-$ -0.5 dex, which provides less opacity. This results in an effective temperature that is about 80 K hotter than the layered profile, at $\sim$940 K.

\begin{figure}[h!]
    \centering
    \includegraphics[width=0.99\textwidth]{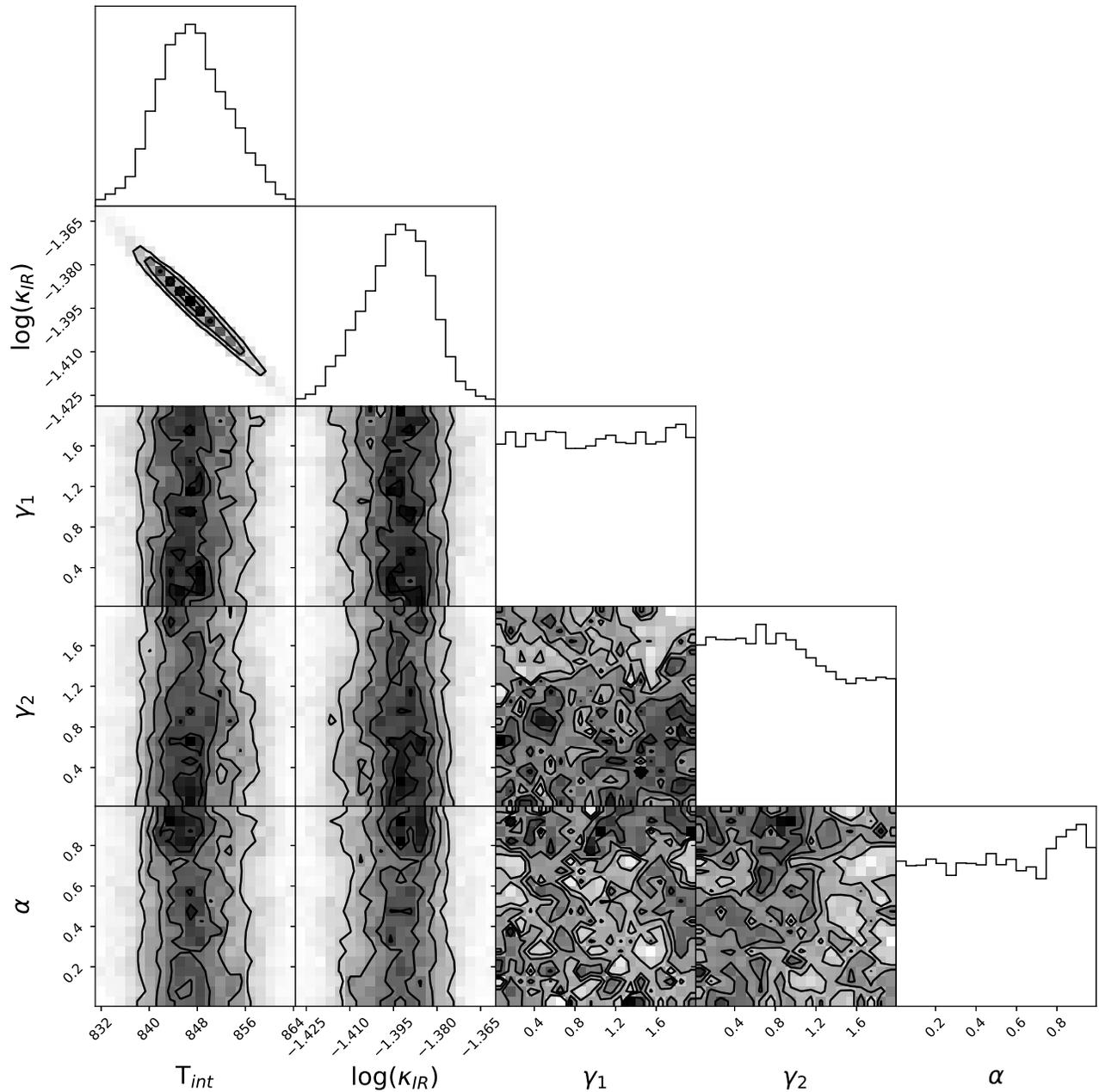}
    \caption{Posterior distributions for the parameters of the best fit parametric pressure profile for GJ 229B.}
    \label{fig:paramcorner}
\end{figure}

In addition to the effective temperature, the parametric T-P profile shown in Figure \ref{fig:GJ229B_TP} exhibits significant differences from the layered one. The parametric profile has higher temperatures above the $\sim$1 bar level with an isothermal upper atmosphere above the $\sim$300 mbar level. The layered profile has a small temperature gradient at these altitudes, but a nonzero one.

We also note the unusual kink in the layered profile at the $\sim$10 bar level. This feature is tightly constrained and occurs in a region of the atmosphere from which a large fraction of the flux originates. While it has not been previously observed in retrievals of comparable objects (e.g. \citealt{Line2015}), it appeared consistently in our layer-by-layer fits for this spectrum under various test conditions. The origin of this feature in the temperature-pressure profile is unclear. The most natural explanation would involve the presence of an absorber occurring at that altitude, which partially blocks the outgoing thermal radiation and disproportionately heats that layer. This absorber could be a non-equilibrium molecular component or an aerosol of a condensate such as Na$_2$S (see e.g. \citealt{Sing}). Further research on the likely atmospheric chemistry of brown dwarfs is needed to identify potential candidates for such an absorber.


The posterior distribution of the five parameters of the T-P profile is shown in Figure \ref{fig:paramcorner}. Almost all of the variation in the profile is determined by the first two parameters, the intrinsic temperature (effective temperature for non-irradiated objects) and the mean infrared opacity. Given a fixed radius and surface gravity, these parameters are also tightly correlated with each other, but they continue to determine almost all of the variation in either case.


\subsection{Bulk Parameters and Evolutionary Model Comparison}
\label{sec:evofits}


An interesting feature of these retrievals is that in all cases, the surface gravity is tightly constrained -- tightly enough that the consistency between different retrievals is not very good.

There is still good agreement in specific cases. For the two retrievals of the full spectrum with a layered T-P profile where gravity was a free parameter, they were consistent. Likewise, for the two retrievals with a parameter profile and gravity as a free parameter, they are consistent. This suggests that APOLLO is doing a good job of resolving the degeneracies in the bulk parameters of radius, gravity, and effective temperature. If these degeneracies were poorly constrained, the retrieved values would likely be spread much farther apart in parameter space.

On the other hand, the retrievals of surface gravity between the full spectrum and the various limited data sets are not consistent. This suggests that there are other degeracies related to the wavelength coverage of the spectrum. Surface gravity can be degenerate with molecular abundances because it affects the column density and pressure broadening of absorbers for a fixed volumetric density. As is discussed in Section \ref{sec:chem}, we do observe a significant degeneracy between gravity and the abundances of the major molecular species in GJ 229B. With different wavelength ranges covering different molecular absorption bands, each retrieval is likely to settle on a different position on the correlation curve based on the dominant species in that wavelength range. In this case, a retrieval on the full spectrum, which covers as many bands as possible, is likely to give the most accurate measurement of the true surface gravity.

\begin{figure}[h!]
    \centering
    \includegraphics[width=0.99\textwidth]{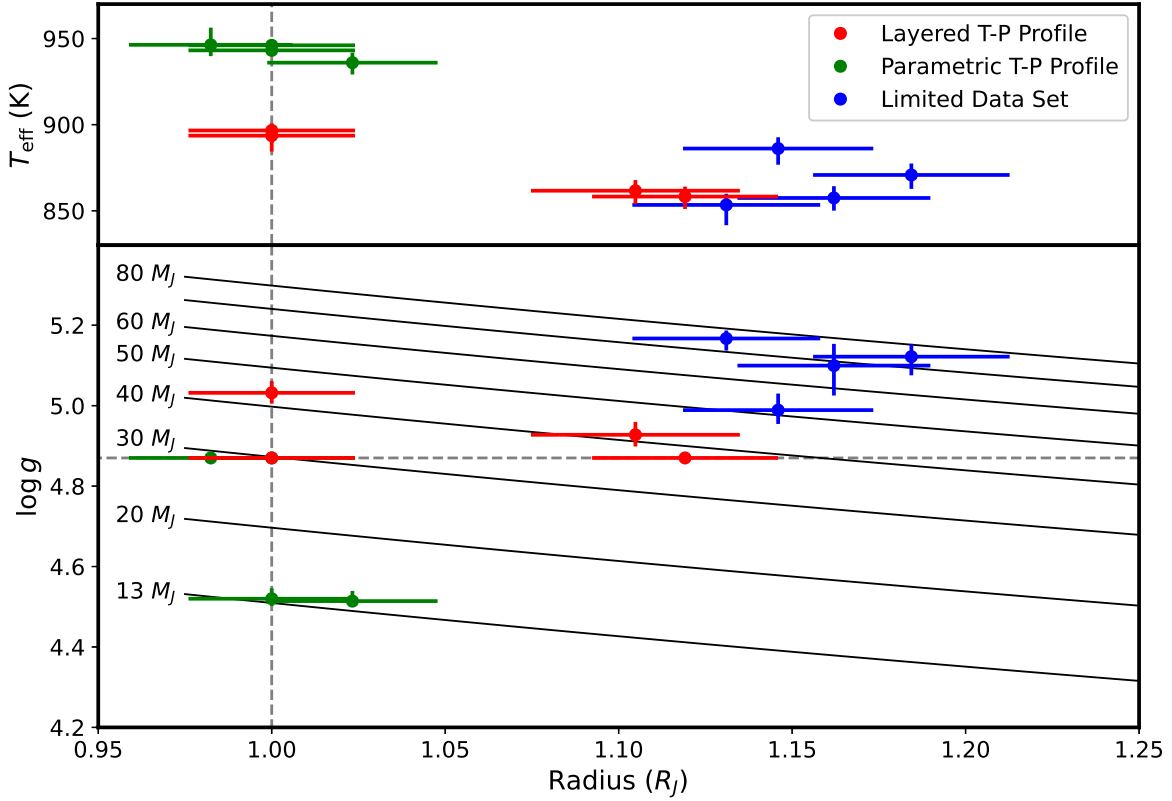}
    \caption{Retrieved bulk parameters for GJ 229B for each of our retrievals. Top panel: effective temperature versus radius. Bottom panel: surface gravity versus radius on the same scale. Models using the full data set and a layered T-P profile are marked in red. Models with limited data sets are marked in blue, and models with parametric T-P profiles are marked in green. Models with a fixed radius lie along the vertical dashed line, and models with a fixed surface gravity lie along the horizontal dashed line. Diagonal black lines are lines of constant mass. \\ {\bf Takeaway:} retrievals with layered T-P profiles, parametric T-P profiles, and limited data sets produce similar results within each group, but diverge significantly between the groups.}
    \label{fig:GJ229B_RM}
\end{figure}

In Figure \ref{fig:GJ229B_RM}, we plot the best-fit values for T$_{\rm eff}$, radius and log-g for all of our retrievals. In this figure, we designate the three groups of retrievals by color: the full data set with a layered T-P profile in red, the limited data sets in blue, and the parametric T-P profiles in green. The diagonal black lines in the bottom panel are lines of constant mass based on radius and surface gravity. The retrievals with fixed radii and/or surface gravities lie on the gray dashed lines.

\begin{figure}[h!]
    \centering
    \includegraphics[width=0.99\textwidth]{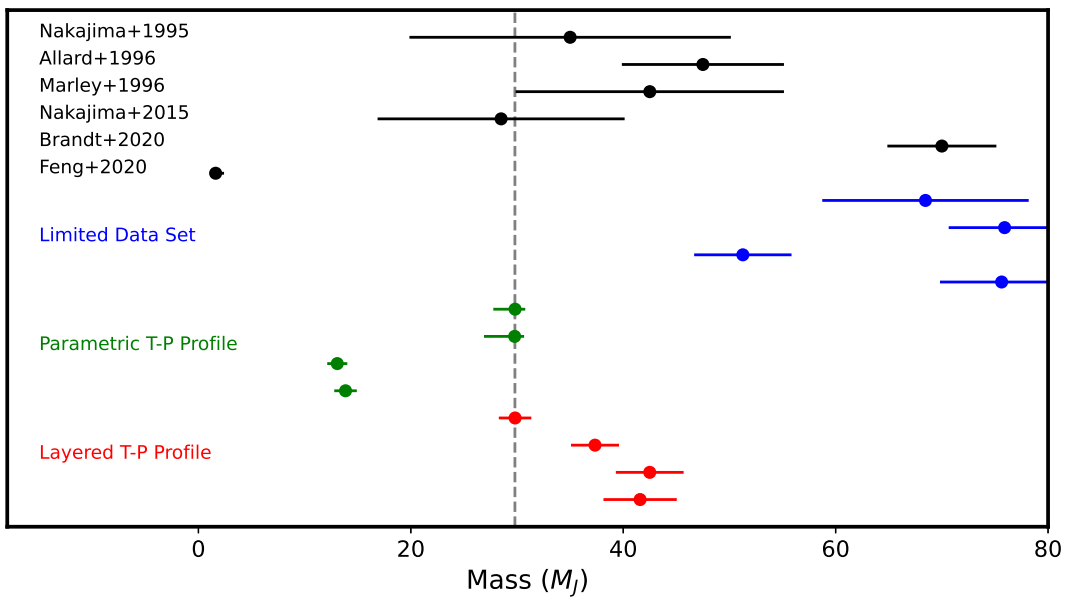}
    \caption{Published mass estimates for GJ 229B compared with our retrieved values for each of our 12 retrievals. Note that the estimate of \cite{GJ229Feng} is an $M{\rm sin}\,i$ value. The vertical dashed line is the fixed mass set by our retrievals with fixed radius and surface gravity.}
    \label{fig:GJ229B_mass}
\end{figure}

For a clearer comparison of our retrieved mass values, we plot the retrieved masses from all our retrievals in Figure \ref{fig:GJ229B_mass} (note that the estimate of \cite{GJ229Feng} is an $M{\rm sin}\,i$ value; we are therefore implicitly assuming an edge-on inclination). Our two retrievals with both fixed radius and fixed surface gravity have a mass set to 29.8 $M_J$, which is indicated by the vertical dashed line. Except for the two recent dynamical estimates, all of the published estimates are (narrowly) in agreement with one another, and our adopted mass of $41.6\pm3.3\, M_J$ is at least marginally consistent with all of them (as are the other retrievals on the full dataset using a layered T-P profile).

\begin{figure}[h!]
    \centering
    \includegraphics[width=0.99\textwidth]{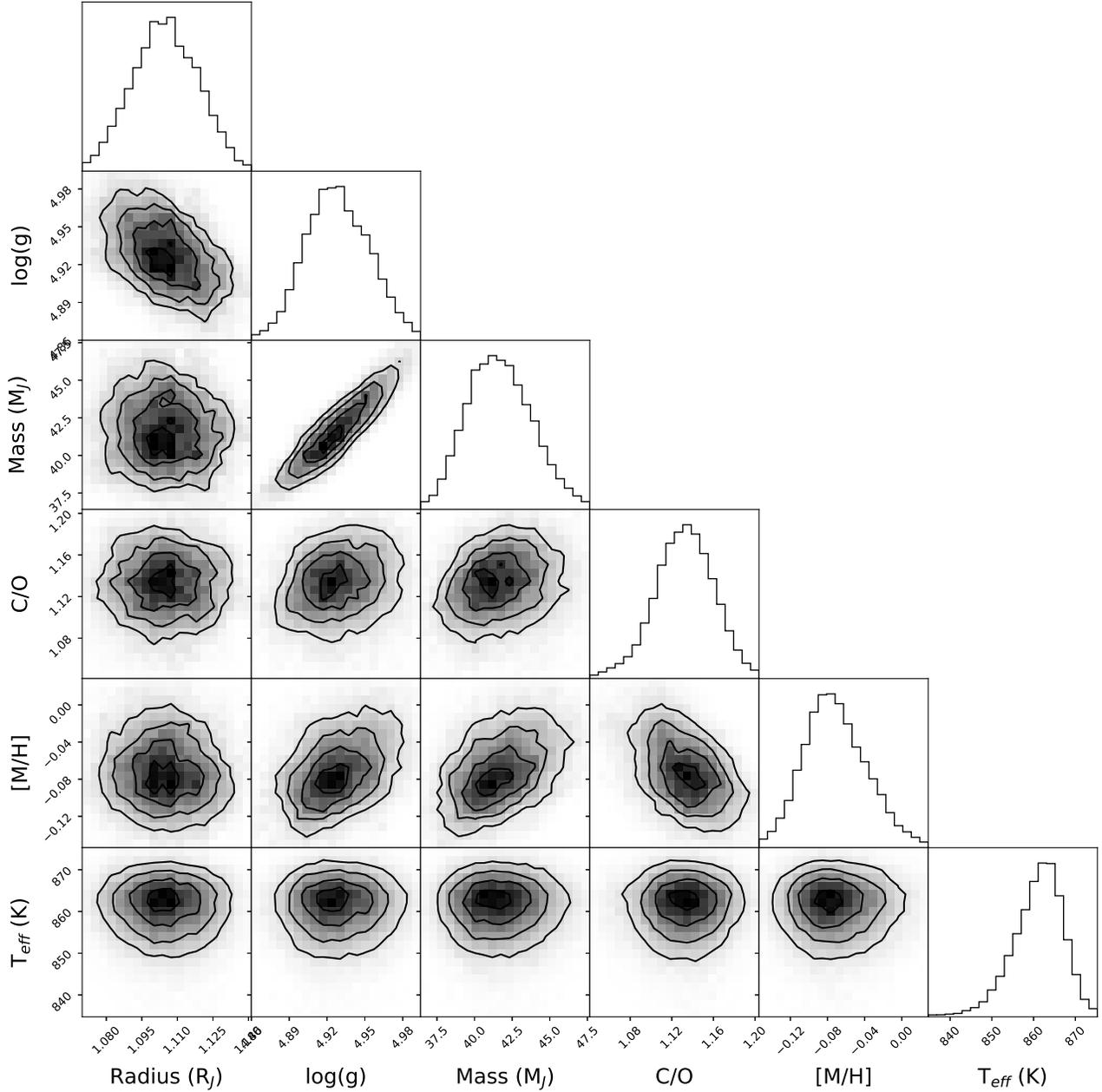}
    \caption{Posterior distributions for the retrieved bulk parameters and derived parameters in our best-fit retrieval of GJ 229B, using the full data set and a layered temperature-pressure profile. Radius and surface gravity are retrieved directly, while the mass, effective temperature, metallicity, and C/O ratio are all derived from retrieved parameters.}
    \label{fig:GJ229B_basic}
\end{figure}

The posterior distributions of the bulk parameters in our best fit retrieval are shown in Figure \ref{fig:GJ229B_basic}. The effective temperature was $869_{-7}^{+5}$ K. As shown by Figure \ref{fig:GJ229B_RM}, the other layered profile retrievals returned similar properties. We note that the limited data set retrievals all returned much higher masses of 50-75 $M_J$, primarily due to a higher surface gravity, while their effective temperatures were similar to the full data set layered profiles. In contrast, the retrievals using a parametric T-P profile, in the cases where the surface gravity was allowed to vary, returned a higher effective temperature with a lower surface gravity, which would imply an unlikely lower value for the mass (near the deuterium burning limit). 

\begin{figure}[h!]
    \centering
    \includegraphics[width=0.9\textwidth]{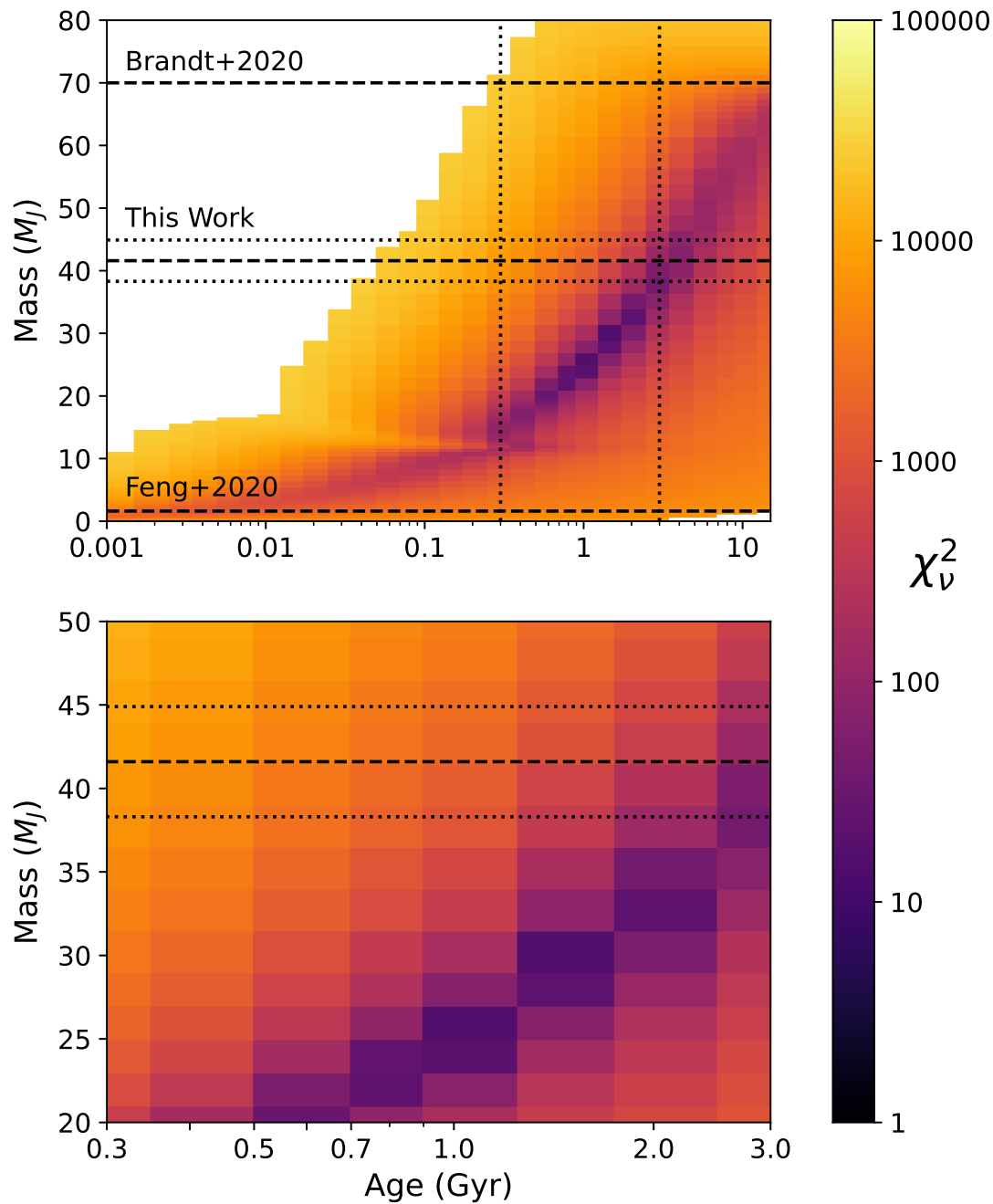}
    \caption{Goodness of fit of the Sonora-Bobcat (S-B) brown dwarf evolutionary models to GJ 229B as a function of mass and age. For each point, a reduced chi-squared statistic was computed based on the goodness of fit to our retrieved values of radius, surface gravity, and effective temperature. The smallest value occurs for $M=30\,M_J$ and $t=1.5$ Gyr, which has $\chi_\nu^2=14.8$. The white regions are not included in the S-B model set. Horizontal dashed line mark publish mass estimates, including this work. Vertical dotted lines mark the published age estimate of \cite{GJ229Prop}.}
    \label{fig:GJ229B_heat}
\end{figure}

\begin{figure}[h!]
    \centering
    \includegraphics[width=0.75\textwidth]{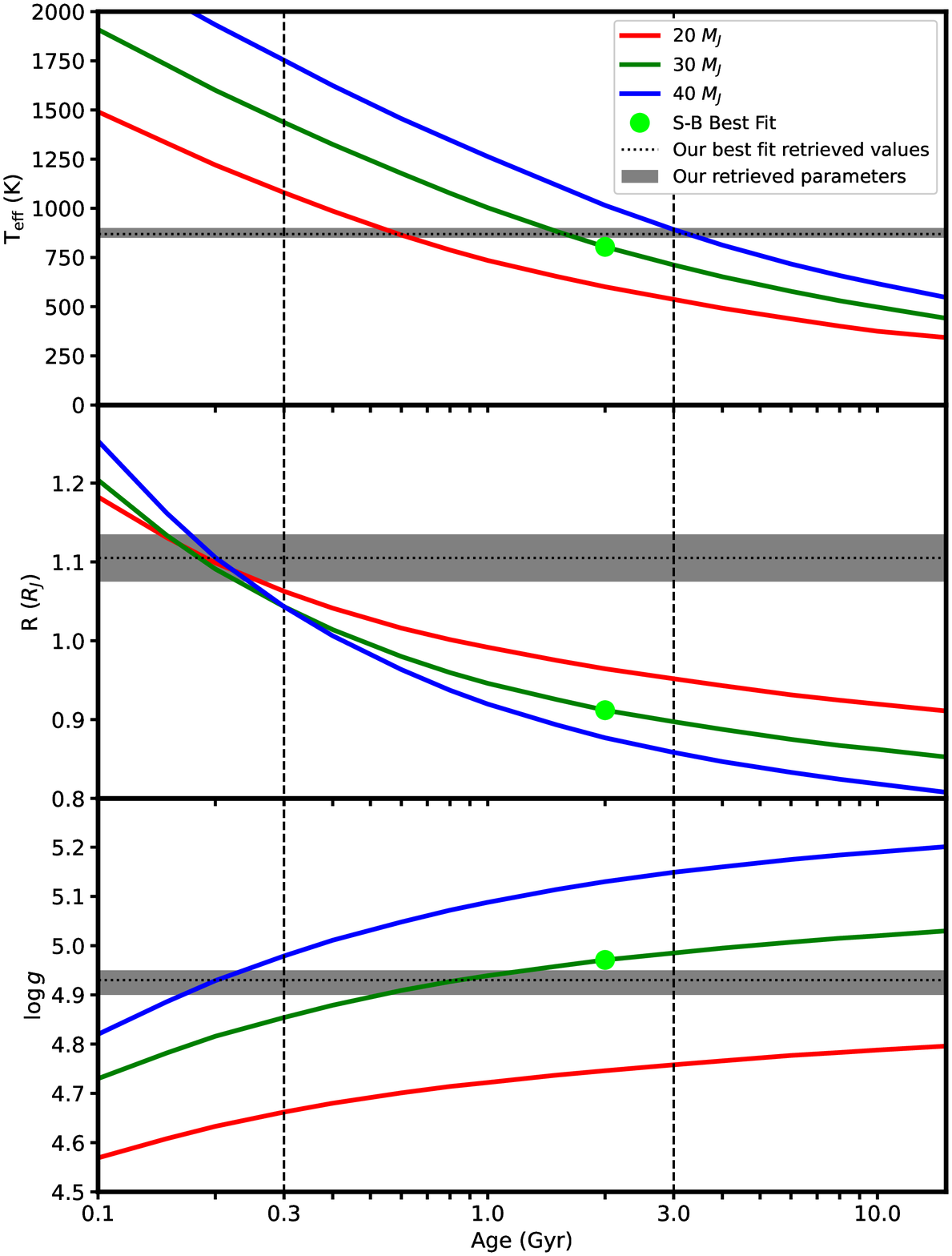}
    \caption{Comparison of Sonora-Bobcat (S-B) brown dwarf evolutionary models with our retrieved parameters of T$_{\rm eff}$, $R$, and ${\rm log}\,g$ (gray bands) for GJ 229B. These three parameters are used to compute the reduced chi-squared values of the S-B models at a given age. The green dot marks the best S-B fit to our retrieved parameters.}
    \label{fig:GJ229B_time}
\end{figure}

To compare these results with the theory of brown dwarf evolution, we computed the goodness of fit of the Sonora-Bobcat (S-B) evolutionary models to our retrieved parameters. The S-B models are presented as a grid in mass and age, with each point having a distinct radius, surface gravity, and effective temperature. To quantify the closeness of each grid point to our retrieved values, we compute the reduced chi-squared statistic between them in these three parameters. A heat map of the reduced chi-squared values for the entire S-B model grid is shown in Figure \ref{fig:GJ229B_heat}. For comparison, the recent dynamical mass estimates for GJ 229B \citep{GJ229Brandt,GJ229Feng} and our retrieved best-fit mass are plotted as horizontal dashed lines. The vertical dotted lines indicate the previous age estimate for GJ 229B of 0.3-3.0 Gyr \citep{GJ229Prop}.

The reduced chi-squared values confirm the very poor fit of the published dynamical mass estimates to our retrieved parameters. Meanwhile, the best fit of any of the S-B models occurs for a mass of 30 $M_J$ and an age of 1.5 Gyr, which corresponds to $\chi_\nu^2=14.8$. This model has bulk parameters of $R=0.962\,R_J$, ${\rm log}g=4.958$, and $T{\rm eff}=883$ K, giving it a significantly smaller radius and modestly higher effective temperature than our retrieved results.

We plot selected S-B evolutionary model tracks in effective temperature, radius, and surface gravity in Figure \ref{fig:GJ229B_time}. {Our best fit retrieved values and uncertainties  for these three parameters are shown as gray bands.} The vertical dashed lines mark the 0.3-3.0 estimated age range, within which the evolutionary tracks should cross through our retrieved parameters. We plot the best-fit S-B mass of 30 $M_J$ (green), a close approximation to our retrieved mass, 40 $M_J$ (blue), and a lower mass model of 20 $M_J$ (red) for comparison. All three of these tracks are consistent with our retrieved parameters within the age limits for effective temperature and radius, and the central 30 $M_J$ track is consistent with our retrieved surface gravity. However, none of them is consistent with our retrieved radius, being 10\%-20\% smaller.

This result reflects a general trend that the S-B models are systematically more compact than our retrieved parameters for the same mass and temperature. For example, the best fit in our retrieved mass range is a 38 $M_J$ model at an age of 3.0 Gyr, which has bulk parameters of $R=0.865\,R_J$, ${\rm log}g=5.120$, and $T{\rm eff}=855$ K.

Such a large discrepancy in radius between our retrieved value and the best evolutionary model fit to our gravity and temperature values could potentially be an indication of, missing physics in the models (leading to a slower cooling and contraction than predicted), incorrect values for the flux calibration, or possibly a binary companion. 
We note that a difference in photometric magnitudes of +0.3 mag from the published values would result in a smaller radius that is consistent with the evolutionary models. While this difference is much larger than the observational uncertainties of 0.05 mag, two earlier measurements of the photometry of GJ 229B yielded significantly higher magnitudes, especially in the J-band: $J=14.2$ \cite{Matthews96} and $J=14.3$ \cite{Leggett99}. If there are as-yet-unknown systematic errors or variability in the photometry of GJ 229B, this could explain the radius discrepancy.

The availability of age estimates in the S-B models allows us to narrow the previous estimate on the age of the GJ 229 system of 0.3-3.0 Gyr. However, the lack of an exact fit makes it difficult to draw precise conclusions. If we look solely at the best-fit S-B model to our retrieved parameters, this would indicate a clear, singular age estimate of 1.5 Gyr. Alternatively, we can look at the best fits among the S-B models that fall within or at the boundary of our retrieved mass range, namely the 38, 40, and 43 $M_J$ tracks. Along these tracks, the best fits occur at 3.0, 3.0, and 4.0 Gyr, respectively. The poor fit makes it difficult to draw robust conclusions, and if the true radius is smaller than our retrieved value, both the mass and age estimates will decrease. However, these results do point to an older age for the system of $>$1.0 Gyr.

\subsection{Atmospheric Structure and Chemistry}
\label{sec:chem}

\begin{figure}[h!]
    \centering
    \includegraphics[width=0.99\textwidth]{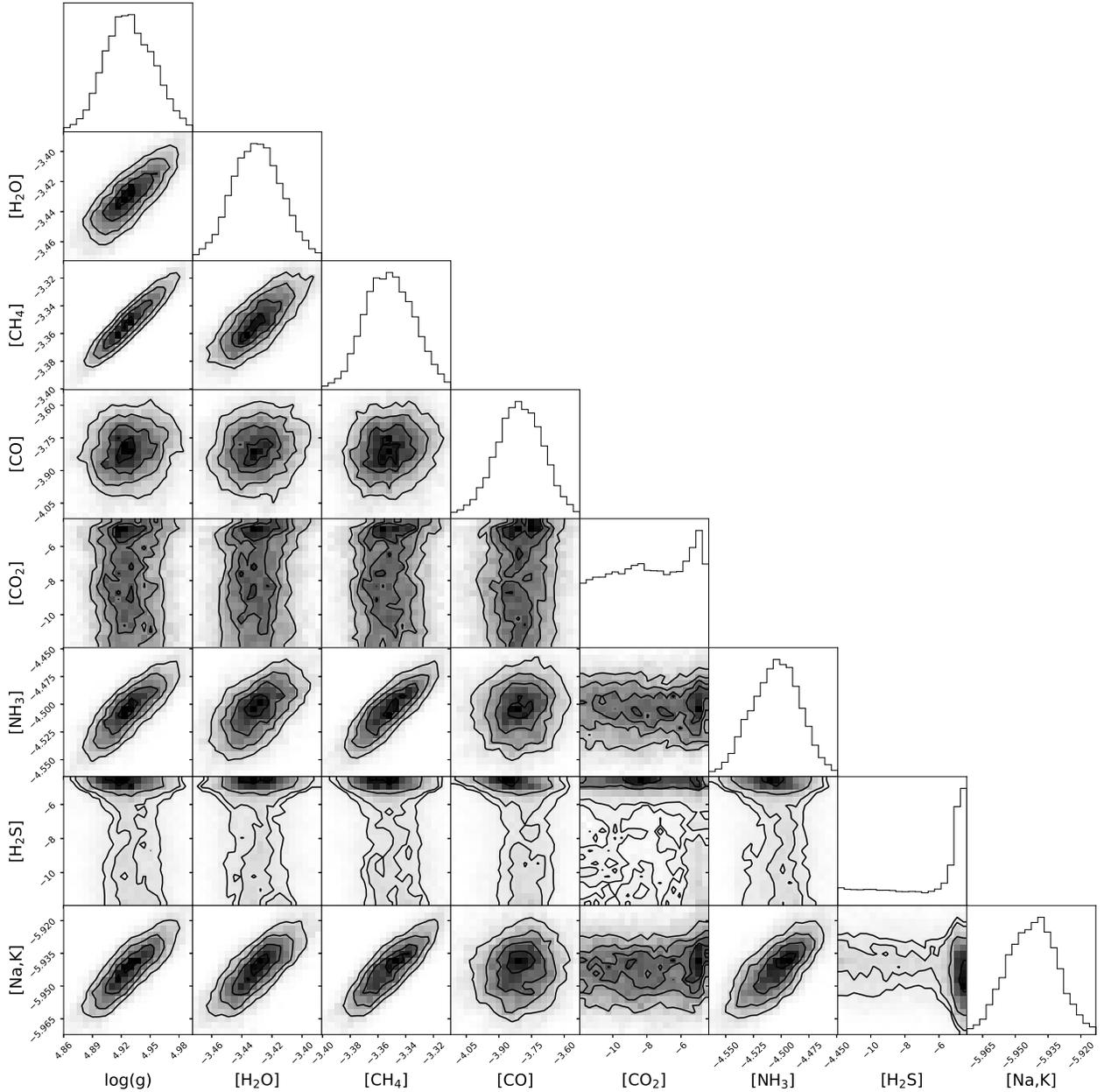}
    \caption{Posterior distributions for the retrieved molecular abundances and surface gravity for our best-fit retrieval of GJ 229B.}
    \label{fig:GJ229B_gas}
\end{figure}

We retrieve abundances for seven common atmospheric constituents; the adopted parameters from the best fit retrieval are listed in Table \ref{tab:best}. We plot the posterior distributions for the molecular abundances plus gravity in Figure \ref{fig:GJ229B_gas}.  We examine the retrieved abundances across the full set of retrievals in Section \ref{sec:riz}.

The posterior distributions indicate significant degeneracies between the surface gravity and the most abundant broad spectrum absorbers, H$_2$O, CH$_4$, and NH$_3$, as well as with alkali metal abundance. Notably, there is negligible degeneracy between surface gravity and CO abundance, which has narrower absorption bands mostly confined to the M-band.

The primary constituents of the atmosphere other than hydrogen and helium are H$_2$O and CH$_4$, followed closely by CO. Notably, we do see a high CO abundance compared with other T dwarf retrievals, which show best fit values of [CO]$\,\sim-7.5$ and even upper bounds of [CO]$\,\sim-5$ for slightly cooler T8 dwarfs (e.g. \citealt{Line2015}). The absolute abundance of CO we measure is $\sim$35\% as great as that of CH$_4$, while the next most abundant molecule, NH$_3$ is only one fifth as common as CO. Our retrieved abundance of [CO]$\,=-3.81_{-0.11}^{+0.12}$ is consistent with prior observations that measured it by fitting individual molecular bands -  \cite{GJ229UKred} reported a CO abundance of $>$50 ppm, or [CO]$\,>-4.30$, while \cite{GJ229Keck} reported an abundance of 200 ppm, or [CO]$\,>-3.70$.

Similarly, \cite{Saumon00} fit three forward models with both CO and NH$_3$. The model most consistent with our retrieved effective temperature gave [CO]$\,=-4.0\pm0.25$. However, they found a much lower NH$_3$ abundance of [NH$_3$]$\,=-5.55\pm0.15$. Our retrieved value of [NH$_3$]$\,=-4.50\pm0.02$ is, significantly higher. This may be due to the overall metal-poor composition assumed by \cite{Saumon00}. However, our result is consistent with that of the T8 dwarfs GJ 570D and HD 3651B \citep{Line2015}.

Our retrieved metallicity is close to Solar at [M/H]$\,=-0.07_{-0.03}^{+0.04}$. This is close to but slightly lower than the most recent grid-based estimate \citep{GJ229Prop}, which proposed [M/H]=+0.13.

\section{Discussion}
\label{sec:discuss}

These results provide us with more precise information about the properties of GJ 229B than was previously available. They also suggest optimal observational and statistical methods for studying similar objects, including both brown dwarfs and young giant planets.

\subsection{Forward Model Design for Retrievals of T-Dwarfs}

The clearest result from our set of retrievals is that a layer-by-layer temperature-pressure profile achieved a significantly better fit to the spectrum than a parametric profile. The primary difference between these two parameterizations is that the layered profile allows more variation in temperature with altitude and in particular allows a non-isothermal upper atmosphere, while the parametric profile we use approximates this region as isothermal. With the largest share of the flux coming from the lower atmosphere, this suggests that absorption by constituents of the upper atmosphere with a non-isothermal temperature gradient significantly sculpts the emergent spectrum so that the more complex profile is needed to capture the effect. This illustrates the need to include the necessary degrees of freedom in the forward model to accurately fit the spectrum.

Other physical effects could potentially cause the same behavior, most notably variation in molecular abundances with altitude due to changes in chemistry or non-equilibrium effects. Given the challenges of performing retrievals with very large parameter sets, this is rarely considered for general retrievals like the ones presented in this paper, as each species would need its own profile. In such cases, metallicity combined with an equilibrium chemistry model is frequently used instead.

For data of the quality available for GJ 229B, the constant molecular abundances we use are adequate to obtain accurate results. However, for to fit higher-resolution and higher-precision data such as those expected from \textit{JWST} and ELTs, further research is needed, perhaps combining equilibrium and non-equilibrium chemistry models with bulk elemental abundances, to determine the most useful degrees of freedom for this purpose. Conversely, because different wavelengths probe different altitudes, high-resolution retrievals of particular molecular bands could help to constrain this vertical structure and also suggest better priors for wide-band retrievals (e.g. \citealt{BrogiHiRes}).

\subsection{Limited Data Set Retrievals and Wavelength Selection}
\label{sec:riz}

\begin{figure}[h!]
    \centering
    \includegraphics[width=0.99\textwidth]{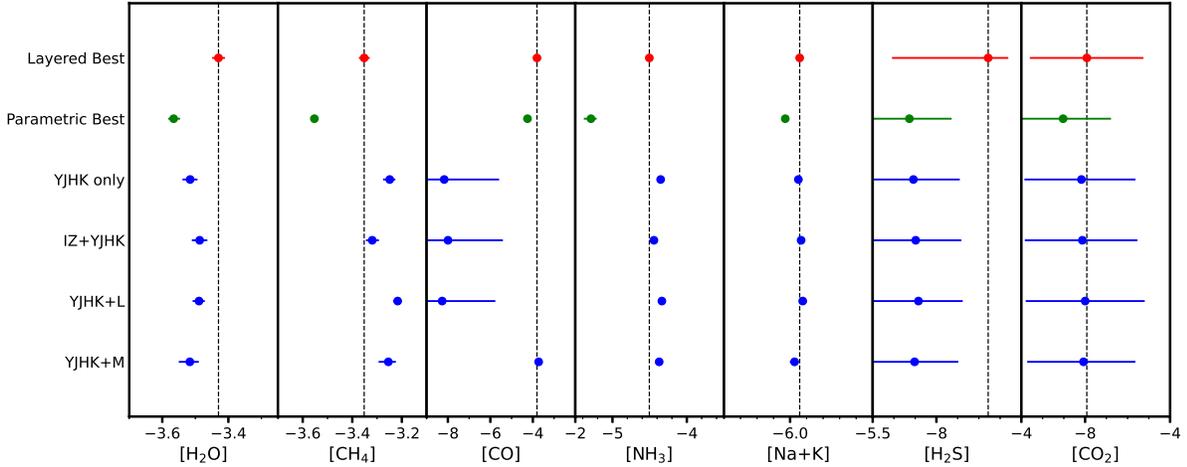}
    \caption{Retrieved molecular abundances for GJ 229B across our set of retrievals.}
    \label{fig:GJ229B_abund}
\end{figure}

Because atmospheric chemistry is central to the models of planet formation we wish to test, it is important to design observations that can accurately retrieve the molecular abundances. The large data set available for GJ 229B allows us to test the usefulness of different wavelength ranges through our limited data set retrievals. We plot the retrieved molecular abundances from our layered and parametric best fit retrievals and all of our limited data set retrievals in Figure \ref{fig:GJ229B_abund}. All of the limited data sets show a closer match than the best fit retrieval using a parametric T-P profile. While significant differences remain, five of the seven molecules have similar retrieved results. A sixth, H$_2$S, is poorly constrained overall and is consistent across all of the retrievals. However, the seventh molecule, CO, is not well constrained by retrievals on limited data sets. Of the four limited data sets, only YJHK$+$M retrieves a CO abundance consistent with our best fit.

Of the twelve retrievals we performed, one of them, the one using the RIZ$+$YJHK limited data set \citep{GJ229HST,GJ229UKblue}, failed to converge to a good fit to the observations as we originally set it up, with large parts of the spectrum not fit at all and many of the retrieved parameters being highly inconsistent with the other retrievals. In an attempt to improve the comparison of atmospheric chemistry, we recalculated this retrieval with narrower priors on the T-P profile. However, this did not significantly improve the goodness of fit.

The inconsistency of this retrieval with the others appeared to be due to an unusual feature of the \cite{GJ229HST} spectrum, which APOLLO was not able to replicate accurately. Specifically, almost all of the observed data points blueward of 0.8 ${\rm \mu m}$ (that is, in the 0.52-0.80 ${\rm \mu m}$ range) are consistent with zero flux. Because our fiducial uncertainties are based on the point-to-point variance in the spectrum, these data points also have uncertainties that are a factor of 3-30 smaller than those in the regions with high flux. Furthermore, because a flat stretch of spectrum can be fit accurately by a flat line, the error bar scaling parameters may fail to capture any underestimation of these uncertainties.

The effect of this flat stretch will be to cause the likelihood function to rate any forward models that produce near zero flux in the 0.52$-$0.80 ${\rm \mu m}$ range as higher likelihood, and with the small uncertainties, it would overweight this wavelength range in the calculation. This creates a local maximum in the probability landscape in which the MCMC algorithm can become stuck. To avoid this problem, we cut off the blue end of this data set at 0.8 ${\rm \mu m}$ to change it to an IZ$+$YJHK spectrum. A retrieval on this data set did produce a good fit to the spectrum, so we adopted this as our short-wavelength limited data set retrieval.

Interestingly, when we ran a retrieval on the full data set cut off at 0.8 ${\rm \mu m}$, it did not improve the goodness of fit. This suggests that for the particular data set available for GJ 229B, the inclusion of the L- and M-bands is a ``tipping point'' at which the flat stretch of spectrum is no longer overweighted enough to cause a poor retrieval fit, and the MCMC algorithm is much more likely to find the global maximum. The consistency of the other eleven retrievals in finding good spectrum fits--and the consistency of the retrieved parameters within each group--is another indicator that APOLLO is producing accurate results.

With this modification to the short-wavelength spectrum, the retrieved abundances of H$_2$O and CH$_4$ appear broadly similar to those of our best fit for all of the limited data sets. However, the CO abundance is retrieved accurately only from spectra that include the M-band, likely because of the strong CO feature at 4.7 ${\rm \mu m}$. For early-to-mid T-dwarfs for which the CO abundance is likely to be higher, obtaining good mid-IR spectra coverage will be even more important for retrieving accurate abundances.

However, while the retrieved abundances are broadly similar, the exact C/O ratio is much more sensitive to small differences in the results. All three of the successful limited data set retrievals returned C/O ratios significantly higher than our best fit value of ${\rm C/O}=1.133$. The YJHK and YJHK$+$L cases both retrieve ${\rm C/O}\approx1.8$, while the IZ$+$YJHK and YJHK$+$M cases both retrieve ${\rm C/O}\approx1.5$.

These results suggests a similar problem to the one presented by the spectroscopic mass determination with limited data sets. Dynamical mass estimates for GJ 229B have yielded values that are strangely inconsistent with each other and also inconsistent or barely consistent with evolutionary models of brown dwarfs. Our adopted mass estimate is much more consistent with evolutionary models, albeit still slightly higher than those models alone would suggest.

However, the retrieved masses varied significantly between different retrievals. For the comparison of the layered and parametric T-P profiles, this can be attributed to the better spectrum fit given by the layered profile, which will naturally produce more accurate results. Yet the results retrieved from our limited data set are more surprising. All of the retrievals on limited data sets returned much higher masses for GJ 229B, which are inconsistent with evolutionary models, while the retrievals on the full data set consistently returned lower masses near 30 $M_J$.

The YJHK data set covers one of the most common spectral ranges for observing brown dwarfs (e.g. \citealt{Burgasser06}), but the results of our retrieval on this spectral range suggest that it many not be sufficient to accurately determine either surface gravities (and therefore masses) or C/O ratios. Adding neither the L band nor the M band individually significantly improved upon the mass retrieval, while including the full data set did. This is further support that that wide-spectrum spectroscopic observations, or at least a wide-spectrum photometric SED, which cover more molecular absorption bands, may be needed to accurately constrain the masses of T-dwarfs, given the degeneracies involved, particularly between abundances and surface gravity. Meanwhile, it appears that accurate determination of C/O ratios also requires coverage of more molecular bands, in this case specifically the mid-IR bands associated with CO and CO$_2$.

The retrieval on the full data set does appear to be more accurate than the results of retrievals on limited data sets, based on several lines of evidence: the more physically plausible mass estimate, the better goodness-of-fit statistics, and the consistency of the retrieved molecular abundances to prior work and similar objects. We successfully reproduced the anomalous CO abundance (attributed to vertical mixing) first noted by \cite{GJ229Keck}, and the other abundances are similar to those found for slightly later T8 dwarfs \citep{Line2015}.

\subsection{Implications for Formation}

The C/O ratio in the atmospheres of brown dwarfs and giant planets is expected to be an indicator of the location and processes of their formation \citep{ObergCO}. Brown dwarfs widely separated from their host stars likely formed separately in the birth cluster and should have near-Solar C/O ratios of $\sim$0.5. Meanwhile, brown dwarfs and planets formed from the disk should have atmospheres enriched in carbon due to rainout of H$_2$O and CO$_2$, with C/O$\sim$1.0.

GJ 229B orbits close enough to GJ 229A to have potentially formed from the disk, and our retrieved C/O ratio for GJ 229B of 1.133 is high enough to suggest a disk origin. However, the much more widely separated brown dwarfs GJ 570D and HD 3651B (which are more likely to have formed with near-solar compositions) also have elevated C/O ratios of 1.09 and 1.22, respectively \citep{Line2015}. This similarity suggests that there may be non-equilibrium chemistry producing high measured C/O ratios in T-dwarfs, even from a near-Solar composition. Retrievals of C/O ratios from a larger sample of brown dwarfs are needed to establish a baseline for comparison to better assess this hypothesis.








\section{Conclusion}
\label{sec:conclude}

We have performed atmospheric retrievals on a wide-wavelength (0.5-5.1 ${\rm \mu m}$) composite spectrum of the brown dwarf GJ 229B to derive individual molecular abundances and a temperature-pressure profile using the APOLLO code. These are the first spectral retrievals for this object incorporating all of the currently available spectrum coverage. Twelve retrievals were performed in total to test the effects of restricting the wavelength range of the input spectrum, degeneracies in bulk parameters such as radius, surface gravity and effective temperature, and two different parameterizations of the temperature-pressure profile.

Our best-fit retrieval was a free retrieval of the full data set with a layer-by-layer T-P profile. The bulk parameters of GJ 229B were generally well-constrained and clustered closely between retrievals of the same group, suggesting that APOLLO is able to resolve the degeneracies between temperature, gravity, and metallicity effectively for an observed spectrum with sufficient wavelength coverage.

These bulk parameters are dependent on an overall scaling factor based on the photometric flux calibration for GJ 229B. Smaller fluxes would decrease our estimated radius, mass, and age for this object, which would be more consistent with evolutionary models. New photometry measurements for GJ 229B would help to resolve this tension, either revising or providing more confidence in the previous values. However, our other retrieved parameters are not affected by the flux calibration, and our results have contributed significantly to our understanding of GJ 229B and the general techniques needed to characterize similar objects:

\begin{itemize}
    \item We measure improved estimates of bulk properties for GJ 229B of $41.6\pm3.3\, M_J$ and $T_{\rm eff}=869_{-7}^{+5}$ K. In contrast to recent dynamical mass estimates, these measurements are consistent with the Sonora-Bobcat evolutionary models \citep{Sonora} and imply an older age for the system of $>$1.0 Gyr.
    \item We measure an estimated metallicity of GJ 229B of [M/H]$\,=-0.074_{-0.030}^{+0.028}$. This is consistent with near-Solar composition models as opposed to other proposed low-metallicity models. Our measured abundances reproduce and refine the previously observed high CO abundance of [CO]$\,=-3.81\pm0.11$, and our measured C/O ratio of $1.133_{-0.029}^{+0.027}$ is consistent with results from other late-T dwarfs.
    \item Fitting the spectrum of the GJ 229B requires sufficient complexity in atmospheric structure to produce a good fit, in that layer-by-layer temperature-pressure profile produced a significantly better spectrum for late T-dwarfs than a simpler parametric model with an isothermal upper atmosphere.
    \item Wide-spectrum wavelength coverage, especially in the mid-IR, is needed to accurately measure the properties of T-dwarfs including surface gravity and molecular abundances. In particular, NIR data alone may overestimate C/O ratios significantly, which may be important to determining formation histories.
\end{itemize}


These retrievals suggest observational coverage and statistical techniques needed to spectroscopically characterize both brown dwarfs and young giant planets accurately. These can help to optimize observing campaigns by \textit{JWST} and other next-generation platforms. Further research can help to improve these requirements by examining more closely new atmospheric physics and the effects of specific wavelength ranges.

\acknowledgements

ARH was supported by an appointment to the NASA Postdoctoral Program at NASA Goddard Space Flight Center, administered by Universities Space Research Association under contract with NASA. AMM acknowledges support from GSFC Sellers Exoplanet Environments Collaboration (SEEC), which is funded in part by the NASA Planetary Science Division’s Internal Scientist Funding Model. This work was partially supported by the GSFC Exoplanets Spectroscopy Technologies (ExoSpec), which is part of the NASA Astrophysics Science Division's Internal Scientist Funding Model. We thank Sandy Leggett for helping us understand the details of the spectroscopic observations of this object. We also thank the anonymous referee for their excellent feedback, which has greatly improved the quality of scientific analysis in this paper.

\bibliography{refs}{}
\bibliographystyle{aasjournal}

\end{document}